\documentclass{aa}

\input{psfig}
\newcommand{\be}{\begin{equation}}
\newcommand{\ee}{\end{equation}}
\newcommand{\bea}{\begin{eqnarray}}
\newcommand{\eea}{\end{eqnarray}}
\begin{document}

\title{Cosmological constraints from clustering properties of galaxy
clusters.}
\author{A. Del Popolo\inst{1,2}, N. Ercan\inst{1} and I. S. Ye\c{s}ilyurt\inst{1}}
\titlerunning{Cosmological constraints from
galaxy clusters.
}
\authorrunning{A. Del Popolo et al.}
\date{}
\institute{
     $1$  Bo$\breve{g}azi$\c{c}i University, Physics Department,
     80815 Bebek, Istanbul, Turkey\\
$^2$ Dipartimento di Matematica, Universit\`{a} Statale di Bergamo,
  via dei Caniana, 2,  24127, Bergamo, ITALY 
}

\maketitle

\begin{abstract}
In this paper, we discuss improvements of the Suto et al. (2000) model, 
in the light of recent theoretical developments (new theoretical mass
functions, a more accurate mass-temperature relation 
and an improved bias model) to predict the clustering properties of galaxy
clusters and to obtain constraints on cosmological parameters. %
%
%
%
We re-derive the two-point correlation function of clusters of galaxies for
OCDM and $\Lambda$CDM cosmological models, and we compare these results with
the observed spatial correlation function for clusters in RASS1 (ROSAT All-Sky Survey 1), and in XBACs (X-RAY Brighest Abell-Type) samples. The comparison shows that the best agreement 
is obtained for the $\Lambda$CDM model with $\Omega_{\mathrm{m}%
}=0.3$. 
The values of the correlation length obtained, ($r_0\simeq 28.2 \pm 5.2
{\rm h^{-1}}$ Mpc for $\Lambda$CDM), are larger than those found in the literature and
 comparable with the results found in Borgani, Plionis \& Kolokotronis
(1999). In order to study the possible dependence of the clustering
properties of the X-ray clusters on the observational characteristics
defining the survey, we calculated the values of the correlation length $r_0$
in the catalogues where we vary the limiting X-ray flux $S_{\mathrm{lim}}$. The
result shows 
an increase of $r_0$ with $L_{\mathrm{lim}}$, 
and correlation lengths that are larger than in previous
papers in literature (e.g Moscardini, Matarrese \& Mo 2001 (hereafter MMM);
Suto et al. 2000). 
These differences 
are due essentially to the different M-T, mass function and bias model used
in this paper. 
Then, we perform a maximum-likelihood analysis by comparing the theoretical
predictions to a set of observational data in the X-ray band (RASS1 Bright
Sample, BCS (Rosat Brightest Cluster Sample), XBACs, REFLEX (ROSAT-ESO Flux Limited X-Ray Sample)), similarly to MMM. In the framework of cold dark
matter models, we compute the constraints on cosmological parameters, such
as the matter density $\Omega_{\mathrm{m}}$, the contribution to density due
to the cosmological constant, $\Omega_{\Lambda}$, the power-spectrum shape
parameter $\Gamma$ and normalization $\sigma_8$. If we fix $\Gamma$ and $%
\sigma_8$, at the values suggested by different observational datasets, we
obtain (for flat cosmological models with varying
cosmological constant $\Omega_{\mathrm{0 \Lambda}} = 1 -\Omega_{\mathrm{0m}}$%
) constraints on the matter density parameter: $0.25 \leq \Omega_{\mathrm{0m}%
} \leq 0.45$ and $0.23 \leq \Omega_{\mathrm{0m}} \leq 0.52$ at the 95.4 and
99.73 per cent levels, respectively, which is 20-30 \% larger than the values
obtained MMM. 
%
%
Leaving $\Gamma$, and $\Omega_{\mathrm{m0}}$, free 
for the flat model, the constraints for $\Gamma$ are $0.1 \leq
\Gamma \leq 0.14$, while for the open model $0.09 \leq \Gamma
\leq 0.13$. These values are smaller than those of MMM by
about $20-30 \%$. If we keep the values of $\Omega_{\Lambda}$ fixed, we obtain
the constraints in the $\Gamma-\sigma_8$ plane. %
%
For the open model with $\Omega_{\mathrm{0m}}=0.3$ the $2\sigma$
region for $\Gamma$ is 0.11-0.2 for $\sigma_8$ it is 0.7 and
1.55. 
For the flat model with $\Omega_{\mathrm{0m}}=0.3$ the $2\sigma$
region has $0.13 \leq\Gamma \leq 0.2$ and $0.8 \leq \sigma_8 \leq 1.6$ The
values of $\sigma_8$ obtained are larger than those of MMM by $\simeq 20 \%$%
. %
If we allow the shape parameter to vary, we find that the clustering
properties of clusters are almost independent of the matter density
parameter and of the presence of a cosmological constant, while they appear
to be strongly dependent on the shape parameter. 

\end{abstract}


\titlerunning{Cosmological constraints from
galaxy clusters.
} \authorrunning{A. Del Popolo \& N. Ercan}

\offprints{A. Del Popolo, E-mail:antonino.delpopolo@boun.edu.tr} 
\institute{
$1$  Bo$\breve{g}azi$\c{c}i University, Physics Department,
     80815 Bebek, Istanbul, Turkey\\
$^2$  Universit\`{a} Statale di Bergamo, Dipartimento di Matematica,
  via dei Caniana, 2,  24127, Bergamo, ITALY \\
}


\begin{keywords}
cosmology: theory - large scale structure of universe - galaxies: formation
\end{keywords}

\section{Introduction}

X-ray studies of clusters of galaxies have provided a large amount of
quantitative data for the study of cosmology. The mass of a rich cluster is
approximately $10^{15} {\rm h^{-1}M_{\odot }}$. \footnote{$%
h={\rm H_{0}/(100kms^{-1}Mpc^{-1})}$, ${\rm H_{0}}$ being the Hubble constant at the
current epoch (in the paper we adopt ${\rm h=0.65}$) (see Spergel et al. (2003),
Tegmark et al. (2004)).} This mass lies within a region of diameter $\simeq
20{\rm h^{-1}Mpc}$ and consequently the observations of clusters can provide
information on the mass distribution of the Universe on these scales.
Furthermore, since rich clusters are rare objects, their properties are
expected to be sensitive to the underlying mass density field from which
they originated. Therefore, clusters of galaxies appear to be ideal tools
for testing theories of structure formation as well as studying large-scale
structure.

X-ray catalogues of X-ray selected clusters are now available from ROSAT:
RASS1 (De Grandi et al. 1999) (ROSAT All-Sky survey 1), BCS (Ebeling et al.
1998) (ROSAT Brightest Cluster Sample), XBACs (Ebeling et al. 1996) (X-ray
brightest Abell Cluster Sample), REFLEX (B\"{o}hringer et al. 1998)
(ROSAT-ESO Flux-Limited X-ray sample) and the volume covered by the samples
is expected to increase through the X-ray satellites such asAstro-E, Chandra,
and XMM. These data together with optical data have been used to compute the
cluster number counts and the X-ray luminosity function, which have relevant
cosmological implications. In particular, the analysis of the cluster
abundance (also as a function of redshift) has been used widely to provide
estimates of the mass fluctuation amplitude and of the matter density
parameter $\Omega _{\mathrm{m}}$, with several, often discrepant results
(Kitayama \& Suto 1997; Mathiesen \& Evrard 1998; Sadat, Blanchard \&
Oukbir 1998; Reichart et al. 1999a,b; Viana \& Liddle 1999; Blanchard,
Bartlett \& Sadat 1998; Eke et al. 1998; Bahcall, Fan \& Cen 1997;
Fan, Bahcall \& Cen 1997; Bahcall \& Fan 1998; Donahue \& Voit 1999;
Borgani et al. 2001). An alternative approach to the abundance of clusters
is based on the study of the spatial distribution of selected clusters. The
standard statistical tools used with this aim are the (spatial and angular)
two-point correlation function and the power-spectrum.

The two-point correlation function is a fundamental statistical test for the
study of the cluster distribution and is relatively straightforward to
determine from observational data. The spatial correlation function of
galaxy clusters provides an important cosmological test, as both the
amplitude of the correlation function and its dependence upon the mean
intercluster separation are determined by the underlying cosmological model.
Like for the cluster abundance, discrepant results have been found for,
e.g., the correlation length (Hauser \& Peebles 1973; Bahcall \& Soneira
1983; Klypin \& Kopylov 1983; Bahcall \& Cen 1992, Bahcall \& West 1992;
Efstathiou et al. 1992, Governato et al. 1999). %
%

As shown in some papers (Eke et al. 1998; Reichart et al. 1999a,b; Donahue
\& Voit 1999; Borgani et al. 2001; Del Popolo 2003), the reasons leading to
the quoted discrepancies are not only connected to the observational data
used, but other unknown systematic effects may be plaguing a large part of
the quoted results (Reichart et al. 1999a,b; Eke et al. 1998; Donahue \&
Voit 1999; Borgani et al. 2001). Systematic effects entering the quoted
analyses are: 1) The inadequate approximation given by the Press-Schechter
relation (e.g., Bryan \& Norman 1998). 2) Inadequacy in the structure
formation as described by the spherical model leading to changes in the
threshold parameter $\delta _{\mathrm{c}}$ (e.g., Governato et al. 1999). 3)
Inadequacy in the M-T relation obtained from the virial theorem (see Voit \&
Donahue 1998; Del Popolo 2002a). 4) Effects of cooling flows. 
5) Missing high redshift clusters in the data used (e.g., the \textit{EMSS}%
). 6) Evolution of the L-T relation.

Although the quoted uncertainties have so far been of minor importance with
respect to the paucity of observational data, a breakthrough is needed in
the quality of the theoretical framework if high-redshift clusters are to
contribute to in the high-precision-era of observational cosmology.

Moreover, the proper comparison of the two-point correlation function with
X-ray data, requires better theoretical predictions which take account of
the selection function of X-ray clusters (Kitayama, Sasaki \& Suto 1998),
the luminosity-and time dependent bias (Mo \& White 1996; Jing 1998;
Moscardini et al. 1998), the light-cone effect (Matarrese et al. 1997;
Matsubara, Suto \& Szapudi 1997; Nakamura, Matsubara \& Suto 1998; Yamamoto
\& Suto 1999) and the redshift-space distortion (Hamilton 1998; Matsubara \&
Suto 1996; Suto et al. 2000; Nishioka \& Yamamoto 1999; Yamamoto, Nishioka
\& Suto 1999; Magira, Jing \& Suto 2000).

The above discussion and recent developments in terms of both theory
(improved relations for the mass function, M-T relation, and bias) and
observation (X-ray data) 
suggest that it would be useful to recalculate the two-point correlation
function and to revisit the constraints on cosmological parameters obtained
until now.

Likely in Del Popolo (2003), in the present paper we are principally
interested in studying the effects of these changes on the values of the
cosmological parameters and in comparing them with previous estimates, and
then in the specific values obtained. For this reason, we made a comparison
of the theoretical results with observations using several samples such as
RASS1 and XBAC, even if it is known that the REFLEX is more precise (small
errorbars). The paper is organized as follows: in Sect. 2, we introduce the
model used. Sect. 3 is devoted to the results and Sect. 4 to discussion and
conclusions.

\section{Theoretical model}

\subsection{Redshift-space distortion}

In order to obtain a theoretical model for the spatial two-point correlation
function of galaxies in different cosmologies, we follow and improve the
paper of Suto et al. (2000) (hereafter S2000). Their model takes proper
account of nonlinear gravitational evolution of mass fluctuations,
redshift-space distortion due to the linear peculiar velocity field and to
finger-of-god effect, cluster abundance and bias evolution on the basis of the
Press -- Schechter theory, and the light-cone effect. 

As previously reported, one of the effects to take into account is the
two-point correlation function distortions due to the peculiar velocity
field. We take into account this redshift-space distortion following Cole et
al. (1994), Magira et al.(2000) and Yamamoto et al. (1999). Assuming that
the bias of the cluster density field relative to the mass density field is
linear and scale-independent, the power spectrum in redshift space is well
approximated by: 
\begin{eqnarray}
  P^{\rm S}_{\rm cl}(k,\mu,z)
= P^{\rm R}_{\rm mass}(k,z) \left[b_{\rm cl}(z)\right]^2~
\left[ {1+\beta(z)\mu^2 \over 1+(k\mu\sigma_v)^2/2}  \right]^2,
\label{eq:Pkcl2d}
\end{eqnarray}
where $P_{\mathrm{mass}}^{\mathrm{R}}(k,z)$ is the mass power spectrum in
real space, $\mu $ the direction cosine of the wavenumber vector and the
line-of-sight of the fiducial observer, and $\beta $ is linear redshift-space
distortion (Kaiser 1987), defined by 
\begin{eqnarray}
  \beta(z)={1\over b_{\rm cl}(z)} {d\ln D_1(z)\over d\ln a(z)},
\label{defbeta}
\end{eqnarray}
%
%
where $b_{\mathrm{cl}}(z)$ is the redshift-dependent bias factor and $%
D_{1}(z)$ is the linear growth factor normalized to be unity at the present
time.


The denominator in Eq. (\ref{eq:Pkcl2d}) takes account of the nonlinear
redshift-space distortion (\textit{finger-of-God}) assuming that the
pair-wise velocity distribution in real space is exponential with the
velocity dispersion of $\sigma _{\mathrm{v}}(z)$. %
%

As in S2000, to calculate $\sigma _{\scriptscriptstyle\mathrm{v}}$ we use
the fitting formula of Mo, Jing \& B\"{o}rner (1997). 
%
Averaging Eq. (\ref{eq:Pkcl2d}) over the angle with respect to the
line-of-sight of the observer one obtains $P_{\mathrm{cl}}^{\mathrm{S}}(k,z)$
similarly to S2000 (Eq. 4-7). 
%
The corresponding two-point correlation function of clusters in redshift
space is computed as 
\begin{equation}
\xi _{\mathrm{cl}}^{\mathrm{S}}(R,z)={\frac{1}{2\pi ^{2}}}\int_{0}^{\infty
}dkk^{2}P_{\mathrm{cl}}^{\mathrm{S}}(k,z)j_{0}(kR),  \label{eq:xiScl}
\end{equation}%
%
%
where $j_{0}(kR)$ is the spherical Bessel function.

\subsection{The evolution of the mass auto-correlation function}

To predict the clustering properties of X-ray clusters, we need a
description of the matter covariance function and its redshift evolution. To
this aim we used the method of Smith et al. (2003), 
which is an improvement of the method of 
Peacock \& Dodds (1994), and Peacock \& Dodds (1996) \footnote{%
Based on Hamilton et al. (1991) original ansatz} for evolving $\xi (r,z)$
into the fully non-linear regime. The authors adopted a new approach to
fitting power spectra, based upon a fusion of the halo model and HKLM
(Hamilton, Kumar, Lu, Matthews) scaling. This approach has been empirically
shown to allow an accurate description of a very wide range of power
spectrum data. Their formula reproduced the scale-free power spectrum data
and also the CDM results of Jing (1998) with an rms error better than 7\%
(see Smith et al. (2003) for more details and their Appendix for the fitting
formula).

The linear power spectrum used in this paper, $P_{\mathrm{L}}\propto
k^{n}T^{2}(k)$, uses the Bardeen et al. (1986) transfer function $T(k)$
(Bardeen et al. (1986)(Eq.~(G3)), and the shape parameter $\Gamma $ is given
by: 
\begin{equation}
\Gamma \ =\ \Omega _{\mathrm{0m}}h\exp (-\Omega _{\mathrm{0b}}-\sqrt{h/0.5}\
\Omega _{\mathrm{0b}}/\Omega _{\mathrm{0m}})\ ,
\end{equation}%
(Sugiyama 1995), where $\Omega _{\mathrm{0m}}$ is the baryonic contribution
to the density parameter. In the part of the paper dealing with the direct
comparison with XBACs and RASS1 data, we consider an open CDM model (OCDM),
with matter density parameter $\Omega _{\mathrm{0m}}=0.3$ and $\sigma
_{8}=0.87$, and a low-density flat CDM model ($\Lambda $CDM), with $\Omega _{%
\mathrm{0m}}=0.3$, and $\sigma _{8}=0.93$ (see e.g. Liddle et al. 1996a,b
and references therein).

In the part of the paper dealing with the maximum-likelihood analysis the
value of $\Gamma$ is allowed to vary in the range 0.05--0.5, while $\Omega_{%
\mathrm{0m}}$ ranges from 0.1 to 1 in the framework of both open and flat
models. The normalizations of the primordial power-spectrum, parameterized
by $\sigma_8$ (the r.m.s. fluctuation amplitude in a sphere of $8 h^{-1}$
Mpc) is allowed to vary in the range $0.2\le \sigma_8 \le 2$. In the
maximum-likelihood analysis the cosmological models considered, are defined
by four parameters: $\Omega_{\mathrm{0m}}$, $\Omega_{\mathrm{0\Lambda}}$, $%
\Gamma$ and $\sigma_8$.


\begin{figure}[tbp]
\caption{The bias factor $b(\protect\nu)$ as a function of $\protect\nu^2$.
The solid line represents the spherical collapse prediction of Mo \& White
(1996), the dotted line the prediction for $b$ obtained from the model of
this paper and the dashed line the ellipsoidal collapse prediction of Sheth
\& Tormen (1999). }
\label{Fig. 1}%
\centerline{\hbox{Fig. 1
\psfig{file=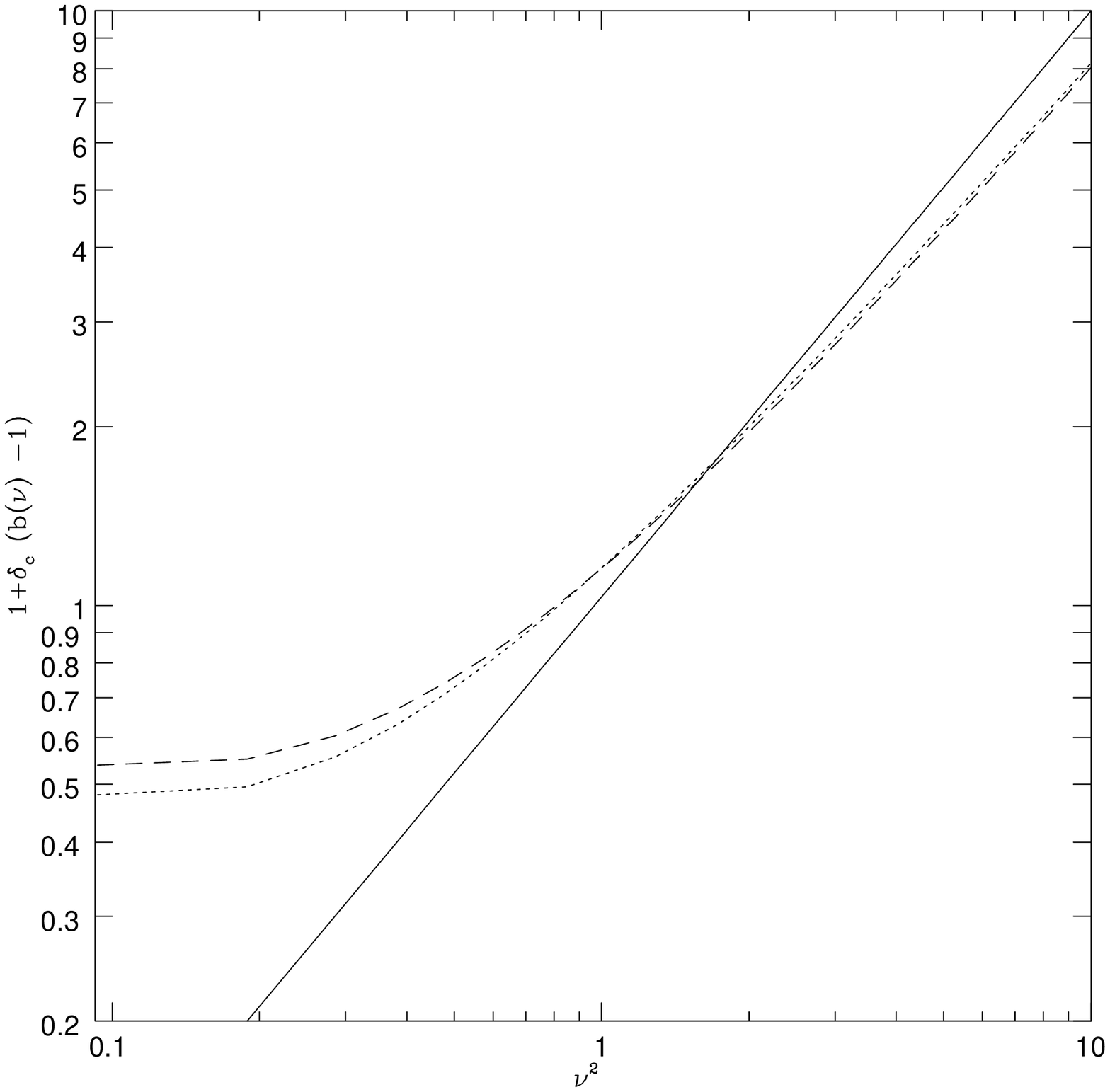,width=7.7cm} Fig. 2
\psfig{figure=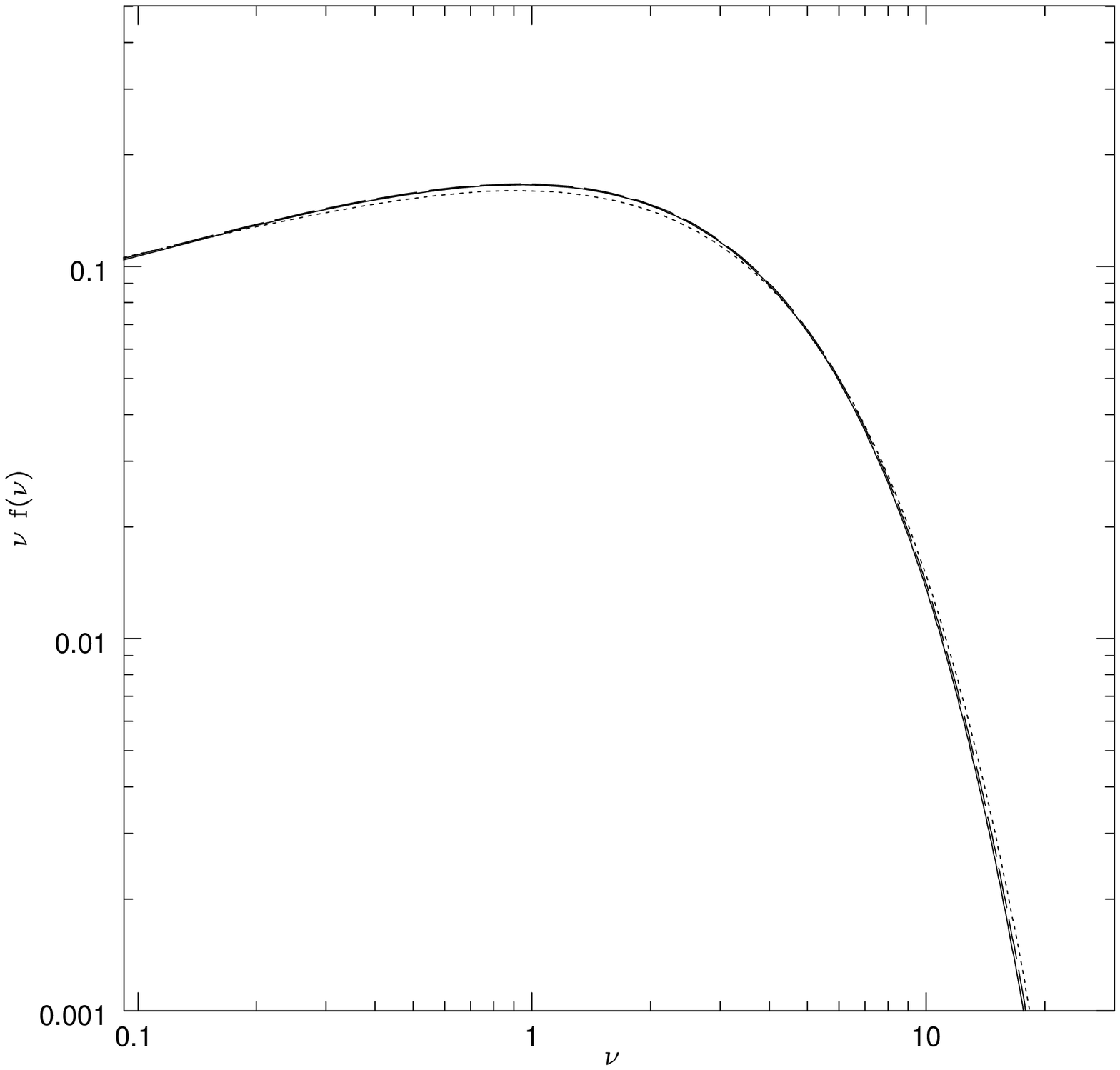,width=7.7cm}
}} \centerline{\hbox{
}}
\caption{Comparison of various mass functions. The dotted line represents the
Sheth \& Tormen (2002) prediction, the solid line that of Jenkins et al.
(2001) and the dashed line that of Del Popolo (2002b).}
\end{figure}

\subsection{Bias evolution}

In order to predict the clustering properties of clusters as a function of
redshift, we need to know how bias evolves. S2000 adopted for the
`monochromatic' bias $b(M,z)$ the expression which holds for virialized dark
matter haloes (see their Eq. 17) (e.g. Mo \& White 1996; Catelan et al.
1998), and to get the effective bias factor the Mo \& White (1996) equation
was combined with the Press-Schechter relation to translate the quoted bias
factor into a function of X-ray flux limit (see their Eq. 18).

Several papers in literature has shown that the Mo \& White (1996) bias formula does not correctly
reproduce the correlation of low mass haloes in numerical simulations.


Several alternative fits have been proposed (Del Popolo \& Gambera 1998;
Jing 1998; Porciani, Catelan \& Lacey 1999; Sheth \& Tormen 1999; Jing 1999;
Del Popolo 2001). %
The bias model of Sheth \& Tormen (1999) 
has been shown to produce an accurate fit of the distribution of the halo
populations in the GIF simulations (Kauffmann et al. 1999). 
\begin{figure}[tbp]
\label{Fig. 1} 
\centerline{\hbox{(a)
\psfig{file=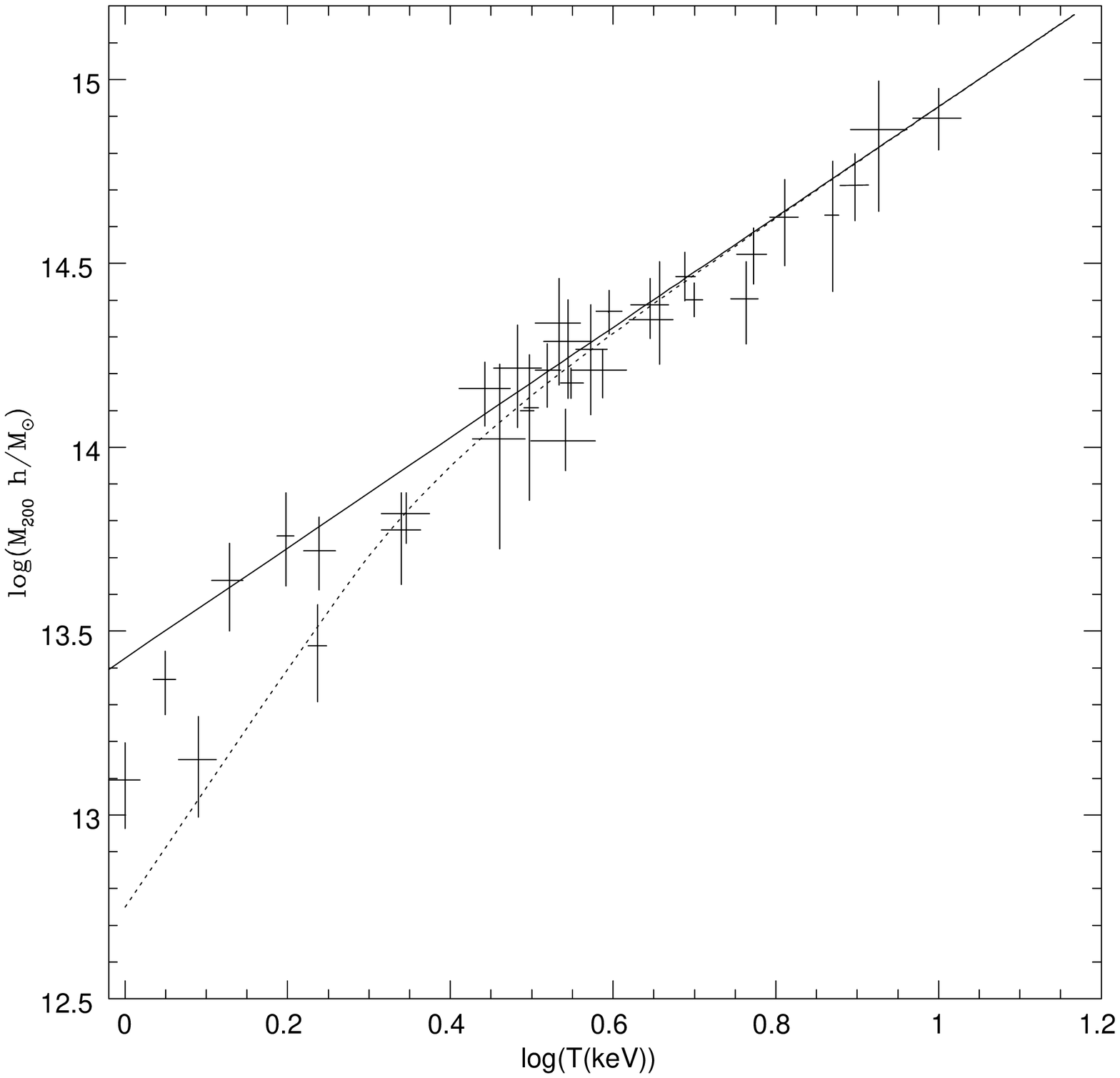,width=8cm} (b)
\psfig{file=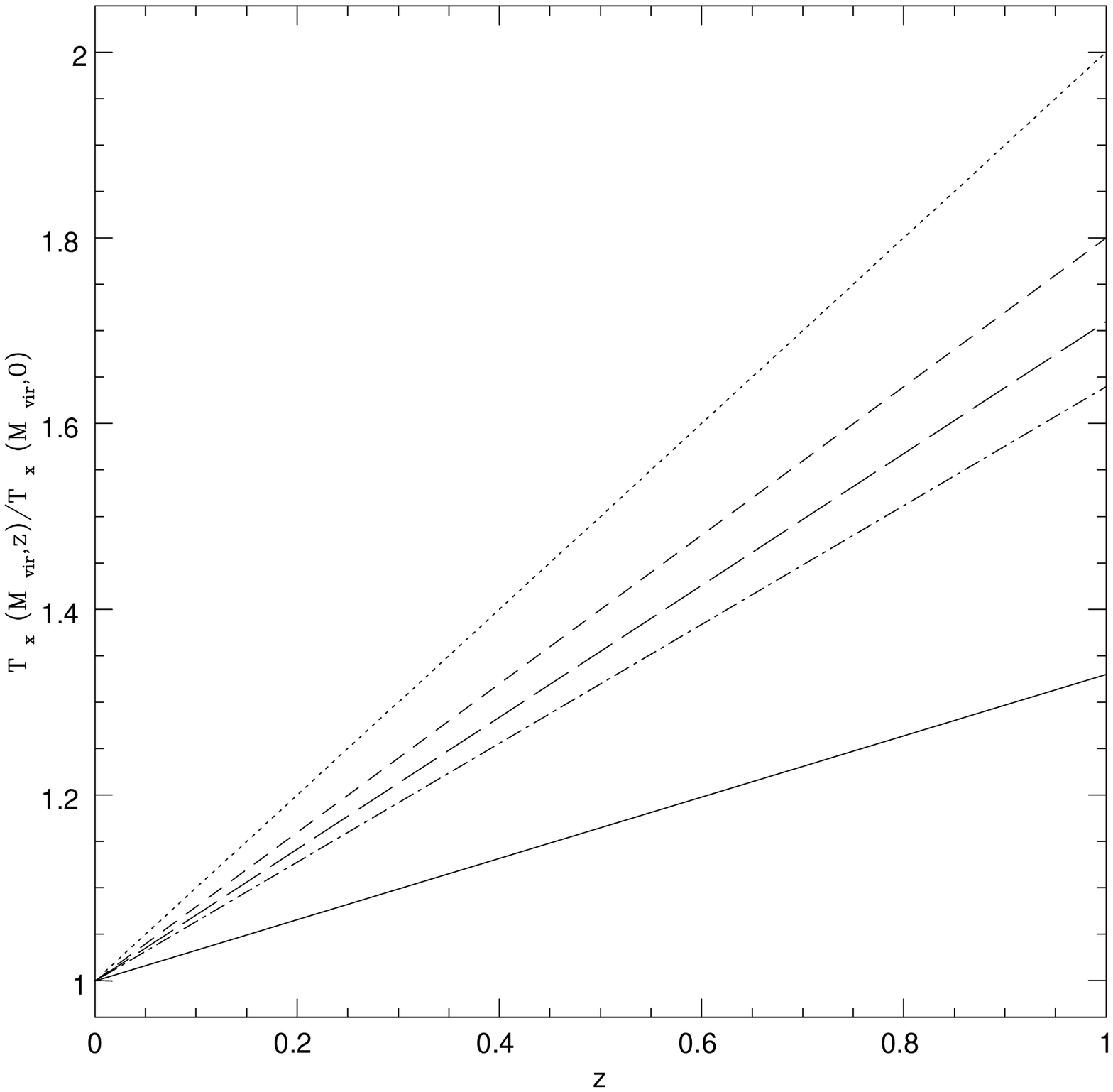,width=8cm}
}}
\caption{(a) M-T relation predicted by the modified continous cluster
formation model. The solid line is the prediction of the quoted
model, described in Del Popolo (2002a), for $\Omega _{\Lambda }=0$, $\Omega
_{\mathrm{0}}=1$, shifted downwards, similarly as in Afshordi \& Cen (2002), to
fit the Finoguenov et al. (2001) observational data at the massive end. The
dotted line represents the prediction of the same model for $\Omega
_{\Lambda }=0.7$ and $\Omega _{\mathrm{0}}=0.3$. (b) Temperature evolution
predicted by the modified continous cluster formation model in the case $%
\Omega _{0}=0.3$. The dotted line represents the ``classical'' prediction, $%
T_{\mathrm{X}}\propto (1+z)$. The short-dashed line represents the
late-formation approximation as expressed by Eq. (8) in Voit (2000), namely $%
T_{\mathrm{X}}\propto \Delta _{\mathrm{vir}}^{1/3}\left[ \frac{\Omega _{%
\mathrm{0}}}{\Omega _{\mathrm{0}}(z)}\right] ^{1/3}(1+z)$. The long-dashed
line and the dot-dashed one plot Eq. (40) in Del Popolo (2002a) for the
spherical collapse model $n=-2$, $n=-1$, respectively. The solid line plots
the same equation for $n=-1$ and taking account of angular momentum
acquisition by the protostructure.}
\end{figure}

In this paper we adopt the bias model described in Del Popolo \& Gambera
(1998), and Del Popolo (2001), 
because it produces a mass function that is in better agreement with the
predictions of Jenkins et al. (2001) (see below). The quoted biased model
is based on the threshold: 
\begin{equation}
\delta _{\mathrm{c}}=\delta _{\mathrm{co}}\left[ 1+\int_{r_{\mathrm{i}}}^{r_{%
\mathrm{ta}}}\frac{r_{\mathrm{ta}}L^{2}\cdot \mathrm{d}r}{GM^{3}r^{3}}\right]
\label{eq:ma7}
\end{equation}%
where $\delta _{\mathrm{co}}=1.68$ is the critical threshold for a spherical
model, $r_{\mathrm{i}}$ is the initial radius, $r_{\mathrm{ta}}$ is the
turn-around radius, and $L$ is the angular momentum. 
\footnote{%
The angular momentum appearing in Eq. ~(\ref{eq:ma7}) is the total angular
momentum acquired by the proto-structure during evolution. In order to
calculate $L$, We will use the same model as described in Del Popolo \&
Gambera (1998, 1999) (more hints on the model and some of the model limits
can be found in Del Popolo, Ercan \& Gambera (2001)).} 
Eq. (\ref{eq:ma7}), is obtained taking account of the effect of
asphericities. In Fig. 1, we plot the bias parameter, $b$, as a function of
the peak height $\nu $, \footnote{$\nu =\left( \frac{\delta _{\mathrm{c}}(z)%
}{\sigma (M)}\right) ^{2}$ is the ratio between the critical overdensity
required for collapse in the spherical model, $\delta _{\mathrm{c}}(z)$, to
the r.m.s. density fluctuation $\sigma (M)$, on the scale $r$ of the initial
size of the object $M$} which is proportional to the halo mass. The solid
line shows the spherical collapse prediction of Mo \& White (1996), the
dotted line the prediction for $b$ obtained from our model, and the dashed
line the ellipsoidal collapse prediction of Sheth \& Tormen (1999). As shown
in the figure, taking account of the effects of asphericity 
produces a change in the dependence of $b$ on $\nu $ in good agreement with
Sheth \& Tormen (1999). 
From Fig. 1 it is evident that at the low mass end the bias relation has an
upturn, meaning that less massive haloes are more strongly clustered than in
the prediction of the spherical collapse model of Mo \& White (1996) and in
agreement with N-body simulations (Jing 1998; Sheth \& Lemson 1999;
Kauffmann et al. 1999). %
%

The above bias factor can be translated into a function of X-ray flux limit
according to: 
\begin{equation}
b_{\mathrm{eff}}(z,>S_{\mathrm{lim}})={\frac{\displaystyle\int_{M_{\mathrm{%
lim}}(S_{\mathrm{lim}})}^{\infty }dM~b(z,M)~n(z,M)}{\displaystyle\int_{M_{%
\mathrm{lim}}}^{\infty }dM~n(z,M)}}  \label{eq:beff} \\
\end{equation}%
where $n(z,M)$ is the number of objects actually present in the catalogue
with redshift in the range z, z+dz and M in the range M, M+dM. One can
estimate $n(z,M)$ from the Press-Schechter (1974) (PS) formula; 
however, several studies have shown some discrepancies between PS and
simulations.

In order to obtain a better estimate of $n(M,z)$, we can use the excursion
set approach. 
which allows one to calculate good approximations to several important
quantities, such as the ``unconditional" and ``conditional" mass functions.
Sheth \& Tormen (2002) (hereafter ST) provided formulas to calculate these
last quantities starting from the shape of the barrier. They also showed
that the ``unconditional" 
mass function, which is the one we need now, is in good agreement with
results from numerical simulations. Using the barrier shape obtained in Del
Popolo \& Gambera (1998), obtained from the parameterization of the
nonlinear collapse discussed in that paper, together with the results of ST
we can calculate 
the ``unconditional" 
mass function.

Assuming that the barrier is proportional to the threshold for the collapse,
as in ST, the barrier can be expressed in the form: 
\begin{equation}
B(M)=\delta _{\mathrm{c}}=\delta _{\mathrm{co}}\left[ 1+\int_{r_{\mathrm{i}%
}}^{r_{\mathrm{ta}}}\frac{r_{\mathrm{ta}}L^{2}\cdot \mathrm{d}r}{GM^{3}r^{3}}%
\right] \simeq \delta _{\mathrm{co}}\left[ 1+\frac{\beta _{1}}{\nu ^{\alpha
_{1}}}\right]
\end{equation}%
(Del Popolo 2001), where $\delta _{\mathrm{co}}=1.68$ is the critical
threshold for a spherical model, $r_{\mathrm{i}}$ is the initial radius, $r_{%
\mathrm{ta}}$ is the turn-around radius, $L$ the angular momentum, $\alpha
_{1}=0.585$ and $\beta _{1}=0.46$. 

As described in Del Popolo (2002b), the mass function can be approximated
by: 
\begin{equation}
n(M,z)=\frac{\overline{\rho }}{M^{2}}\frac{d\log {\nu }}{d\log M}\nu f(\nu
)\simeq 1.21\frac{\overline{\rho }}{M^{2}}\frac{d\log (\nu )}{d\log M}\left(
1+\frac{0.06}{\left( a\nu \right) ^{0.585}}\right) \sqrt{\frac{a\nu }{2\pi }}%
\exp {\{-a\nu \left[ 1+\frac{0.57}{\left( a\nu \right) ^{0.585}}\right]
^{2}/2\}}  \label{eq:nmm}
\end{equation}%
where $a=0.707$, 
and $\overline{\rho }$ is the background density \footnote{%
Note that in this formula $\nu =\left( \frac{\delta _{\mathrm{c}}(z)}{\sigma
(M)}\right) ^{2}$}. \ In Fig. 2 we plot a comparison of the various mass
functions: the dotted line represents Sheth \& Tormen (2002) prediction, the
solid line that of Jenkins et al. (2001) and the dashed line that of Del
Popolo (2002b). As Fig. 2 shows, the mass function obtained in this paper is
in very good agreement with that of Jenkins et al. (2001) in the regime
probed by the simulations. Notice that a large part of the constraints
obtained in the past has been obtained 
using the PS mass function. Only in some more recent papers has the mass
function been calculated by means of ST model 
(e.g. Borgani et al. 2001) or that of Jenkins et al. (2001) (Hamana et al
2001; MMM). Moreover the 
M-T relation chosen is the usual one obtained simply from the virial theorem
(see next subsection). In other words, this paper introduces noteworthy
improvements on the previous calculations in literature.

Before concluding this subsection, following Hamana et al. (2001), we
further attempt to incorporate the scale-dependence on the basis of the
results of Taruya \& Suto (2000), in which the scale-dependence arises as a
natural consequence of the formation epoch distribution of halos. Yoshikawa
et al. (2001), had shown that the scale-dependence of the Taruya \& Suto
(2000) model agrees with their numerical simulations. 
Therefore we construct an empirical halo bias model of the two-point
statistics which reproduces the scale-dependence of the Taruya \& Suto
(2000) bias with the amplitude fixed by the mass-dependent bias 
obtained in Del Popolo \& Gambera (1998) and Del Popolo (2001). Following
Hamana et al. (2001), 
the scale dependent bias shall be described by the following simple fitting
formula: 
\begin{equation}  \label{eq:bhalo_MRz}
b_{\mathrm{h}}(M,R,z)= b(M,z) \left[ 1.0+ b(M,z)\sigma_R(R,z) \right]^{0.15},
\end{equation}
for $R>2R_{\mathrm{vir}}(M,z)$, and otherwise 0, where $R_{\mathrm{vir}%
}(M,z) $ is the virial radius of the halo of mass $M$ at $z$ and $\sigma_R
(R,z)$ is the mass variance smoothed over the top-hat radius $R$. 
The bias factor $b$ contained in Eq. (\ref{eq:xiScl}) through the spectrum $%
P^{\mathrm{S}}_{\mathrm{cl}}$ should then be substituted by Eq. (\ref%
{eq:bhalo_MRz}).

Although the modeling is not completely self-consistent in the sense that
the scale-dependence of the halo biasing factor is neglected in describing
the redshift distortion, the above prescription provides a good
approximation since the scale-dependence in the biasing is of secondary
importance in the redshift distortion effect of halos (see Hamana et al.
2001). 

%
%
%
%

\subsection{Limiting flux and halo mass}

In order to predict the abundance and clustering of X-ray clusters in a
given sample, it is necessary to relate the X-ray fluxes to the
corresponding halo mass at each redshift. As a first step, we relate the
total mass of the dark halo of a cluster to the temperature of the gas. The
M-T relation that we shall use is that calculated in Del Popolo (2002a). The
M-T relation is calculated using the merging-halo formalism of Lacey \& Cole
(1993), which takes account of the fact that massive clusters accrete matter
quasi-continuously, and the present paper is an improvement of a model proposed by Voit (2000)
(hereafter V2000), again to take account of angular momentum acquisition by
protostructures and of an external pressure term in the virial theorem. The
M-T relation obtained in Del Popolo (2002a), 
is given by: 
\begin{equation}
kT\simeq 8keV\left( \frac{M^{\frac{2}{3}}}{10^{15}h^{-1}M_{\odot }}\right) 
\frac{\left[ \frac{1}{m}+\left( \frac{t_{\Omega }}{t}\right) ^{\frac{2}{3}}+%
\frac{K(m,x)}{M^{8/3}}\right] }{\left[ \frac{1}{m}+\left( \frac{t_{\Omega }}{%
t_{0}}\right) ^{\frac{2}{3}}+\frac{K_{0}(m,x)}{M_{0}^{8/3}}\right] },
\label{eq:kTT1}
\end{equation}%
where $t_{\Omega }=\frac{\pi \Omega _{0m}}{H_{o}\left( 1-\Omega _{0m}-\Omega
_{0\Lambda }\right) ^{\frac{3}{2}}}$, $m=5/(n+3)$ ( where $n$ is the
spectral index), $M_{0}$ is defined in Del Popolo (2002a), and: 
\begin{eqnarray}
K(m,x) &=&Fx\left( m-1\right) \mathit{LerchPhi}(x,1,3m/5+1)-  \nonumber \\
&&F\left( m-1\right) \mathit{LerchPhi}(x,1,3m/5),
\end{eqnarray}%
where $F$ is defined in Del Popolo (2002a) (Eq. 35) 
and the \textit{LerchPhi} function is defined as follows: 
\begin{equation}
LerchPhi(z,a,v)=\sum_{n=0}^{\infty }\frac{z^{n}}{(v+n)^{a}},
\end{equation}%
and where $K_{0}(m,x)$ indicates that $K(m,x)$ must be calculated assuming $%
t=t_{0}$.

Eq. (\ref{eq:kTT1}) takes account of the fact that massive clusters accrete
matter quasi-continuously, and also of tidal interaction between clusters.
The obtained M-T relation is no longer self-similar, there is a break at the
low mass end ($T\sim 3-4\mathrm{keV}$) of the M-T relation is present. The
behavior of the M-T relation is as usual, $M\propto T^{3/2}$, at the high
mass end, and $M\propto T^{\gamma }$, with a value of $\gamma >3/2$
dependening on the chosen cosmology. Larger values of $\gamma $ are related
to open cosmologies, while $\Lambda $CDM cosmologies give a slope
intermediate between the flat case and the open case.

In Fig. (3a), we plot the M-T relation predicted by the modified continous
formation model. The solid line is the prediction of the quoted model,
described in Del Popolo (2002a), for $\Omega _{\Lambda }=0$, $\Omega _{%
\mathrm{0}}=1$, shifted downwards, as in Afshordi \& Cen (2002), to fit the
FRB observational data at the massive end. The dotted line represents the
prediction of the same model 
for $\Omega _{\Lambda }=0.7$ and $\Omega _{\mathrm{0}}=0.3$. 
In Fig. (3b), we plot the temperature evolution predicted by the modified
model for $\Omega _{0}=0.3$. The dotted line represents the ``classical''
prediction, $T_{\mathrm{X}}\propto (1+z)$. The short-dashed line represents
the late-formation approximation as expressed by Eq. (8) in Voit (2000),
namely $T_{\mathrm{X}}\propto \Delta _{\mathrm{vir}}^{1/3}\left[ \frac{%
\Omega _{\mathrm{0}}}{\Omega _{\mathrm{0}}(z)}\right] ^{1/3}(1+z)$. The
long-dashed line and the dot-dashed one plot Eq. (40) in Del Popolo (2002a)
for the spherical collapse model 
$n=-2$, $n=-1$, respectively. The solid line plots the same equation for 
$n=-1$ and taking account of angular momentum acquisition by the
protostructure.

The next step (see S2000) is to transform the temperature to the luminosity
of clusters using the observed luminosity-temperature relation. In S2000,
they assumed: 
\begin{equation}
L_{\mathrm{bol}}=L_{44}\left( \frac{T_{\mathrm{gas}}}{6\mathrm{keV}}\right)
^{\alpha }(1+z)^{\zeta }~~10^{44}h^{-2}\mathrm{erg~sec}^{-1}  \label{eq:lt}
\end{equation}%
%
with $L_{44}=2.9$, $\alpha =3.4$ and $\zeta =0$ on the basis of the quoted
observational indications (e.g., David et al. 1993; Ebeling et al. 1996;
Ponman et al. 1996; Mushotzky \& Scharf 1997).

Several independent analyses of nearby clusters with $T_{\mathrm{X}}\geq 1$
keV consistently show that $L_{44}\simeq 3$ and $\alpha \simeq 2.5-3.5$
(e.g., White, Jones \& Forman 1997; Wu, Xue \& Fang 1999, and references
therein). 
For cooler groups, $\leq $ 1 keV, the $L_{\mathrm{bol}}-T$ relation
steepens. Mushotzky \& Scharf (1997) found that data out to $z\simeq 0.4$
are consistent with no redshift evolution in the $L_{\mathrm{bol}}-T$
relation out to $z\simeq 0.4$. In Moscardini et al. (2000a, 2000b) the
authors translated the cluster bolometric luminosity into a temperature,
adopting the empirical relation 
\begin{equation}
T=\mathcal{A}L_{\mathrm{bol}}^{\mathcal{B}}(1+z)^{-\eta }
\label{eq:luminosrel}
\end{equation}%
where the temperature is expressed in keV and $L_{\mathrm{bol}}$ is in units
of $10^{44}h^{-2}$ erg/s and $\mathcal{A}$ = 4.2 and $\mathcal{B}$ = 1/3,
which are a good representation of the data with $T\geq 1$ keV (e.g.
Markevitch 1998 and references therein). %
%

From what was previously said, it is clear that the L-T relation is a source
of 
uncertainties. As in Del Popolo (2003), in the present paper we are
principally interested in studying the effects of the improvements on the
M-T relation, mass function and bias model on cosmological parameters and to
compare them with previous estimates. 
For this reason, 
in the following, we shall follow the philosophy of Moscardini et al.
(2000a), Borgani, Plionis, \& Kolokotronis (1999), namely we shall adopt a
`default' value for $\alpha $ and $L_{44}$, ($L_{44}=2.9$, and $\alpha =3.4$
as the reference values), and we calculate the correlation function and the
constraints on cosmological parameters. We shall compare these with the
results of previous papers (e.g., Moscardini et al. 2000a, 2000b; S2000).
Finally, we shall calculate the effects of the variation of $\alpha $ (in
the range $2.5\leq \alpha \leq 3.5$) on the resulting model constraints.
Notice that all plots shown in the next sections are based on these
`default' values $L_{44}=2.9$, and $\alpha =3.4$.

%
%
%
After fixing the L-T relation, $L_{\mathrm{bol}}(T_{\mathrm{gas}})$ is
transformed in the band-limited luminosity $L_{\mathrm{band}} [T_{\mathrm{gas%
}},E_1,E_2]$ as shown by S2000 (Sect. 2.2).

To obtain 
$M_{\mathrm{lim}}$, necessary to calculate $b_{\mathrm{eff}}$ in Eq. (\ref%
{eq:beff}), we use the method of S2000 (see their Sect. 2.4) %

\subsection{The light-cone effect}

The final step is to calculate the two-point correlation function on the
light cone (Yamamoto \& Suto 1999): 
\[
\xi _{\mathrm{X-cl}}^{\mathrm{LC}}(R;>S_{\mathrm{lim}})={\frac{{\displaystyle%
\int_{z_{\mathrm{max}}}^{z_{\mathrm{min}}}dz{\frac{dV_{\mathrm{c}}}{dz}}%
~n_{0}^{2}(z)\xi _{\mathrm{cl}}^{\mathrm{S}}(R,z(r);>S_{\mathrm{lim}})}}{{%
\displaystyle\int_{z_{\mathrm{max}}}^{z_{\mathrm{min}}}dz{\frac{dV_{\mathrm{c%
}}}{dz}}~n_{0}^{2}(z)}}} 
\]%
%
where $R$ is the comoving separation of a pair of clusters, $z_{\mathrm{max}%
} $ and $z_{\mathrm{min}}$ denote the redshift range of the survey, and $\xi
_{\mathrm{cl}}^{\mathrm{S}}(R,z;>S_{\mathrm{lim}})$ is the corresponding
two-point correlation function on a constant-time hypersurface at $z$ in
redshift space (Eq.(\ref{eq:xiScl})). The comoving number density of
clusters in the flux-limited survey, $n_{0}(z;>S_{\mathrm{lim}})$, is
computed by integrating the mass function Eq. (\ref{eq:nmm}) as: 
\[
n_{0}(z;>S_{\mathrm{lim}})=\int_{M_{\mathrm{lim}}(S_{\mathrm{lim}})}^{\infty
}n(M,z)dM. 
\]%
%
Finally the comoving volume element per unit solid angle is 
\[
{\frac{dV_{\mathrm{c}}}{dz}}={\frac{d_{\mathrm{C}}^{2}(z)}{H(z)}}. 
\]%
%
where: 
\begin{equation}
H(z)=H_{0}\sqrt{\Omega _{\mathrm{0m}}(1+z)^{3}+(1-\Omega _{\mathrm{0m}%
}-\Omega _{\mathrm{0\Lambda }})(1+z)^{2}+\Omega _{0\Lambda }}  \label{eq:hz}
\end{equation}

\subsection{Maximum-likelihood analysis}

In order to obtain constraints for cosmological models, we use a
maximum-likelihood analysis. One possibility to accomplish the quoted
analysis is as shown by Marshall et al. (1983), Del Popolo (2003), or
Borgani, Plionis \& Kolokotronis (1998).

In the present paper, We 
used the same model as MMM. The likelihood is $\mathcal{L}\propto \exp
(-\chi ^{2}/2)$, where 
\begin{equation}
\chi ^{2}=\sum_{i=1}^{N_{\mathrm{data}}}{\frac{{[r_{0}(i)-r_{0}(i;\Omega _{%
\mathrm{0m}},\Omega _{\mathrm{0\Lambda }},\Gamma ,\sigma _{8})]^{2}}}{{%
\sigma _{r_{0}}^{2}(i)}}}\ .  \label{eq:chi2}
\end{equation}%
The sum runs over the observational dataset described in Sect. 2 of MMM,
i.e. $N_{\mathrm{data}}=3$ and $N_{\mathrm{data}}=4$ for the optical and
X-ray bands, respectively (in the present paper we shall use only X-ray
data). The quantities $r_{0}(i)$ and $\sigma _{r_{0}}(i)$ represent the
values of the correlation length and its $1\sigma $ errorbar for each
catalogue, as reported in Table 1 of MMM; $r_{0}(i;\Omega _{\mathrm{0m}%
},\Omega _{\mathrm{0\Lambda }},\Gamma ,\sigma _{8})$ is the corresponding
theoretical prediction obtained with a given choice of cosmological
parameters. The best-fit cosmological parameters are obtained by maximizing $%
\mathcal{L}$, i.e. by minimizing $\chi ^{2}$. The 95.4 and 99.73 per cent
confidence levels for the parameters are computed by finding the region
corresponding to an increase $\Delta _{\chi ^{2}}$ with respect to the
minimum value of $\chi ^{2}$. Other details of the maximum-likelihood
analysis are given in the next section. 
%
%
\begin{figure}[tbp]
\caption{ Comparison of the cluster space correlations in the RASS1 sample
with the theoretical model of the present paper. In the plot, the dashed line
represents the prediction of Moscardini et al. (2000b) for their $\Lambda $%
CDM model, and the solid line that of the $\Lambda $CDM ($\Omega _{\mathrm{0m%
}}=0.3$, $\protect\sigma _{8}=0.93$) calculated using the model of this
paper.}%
\centerline{\hbox{Fig.4 
\psfig{file=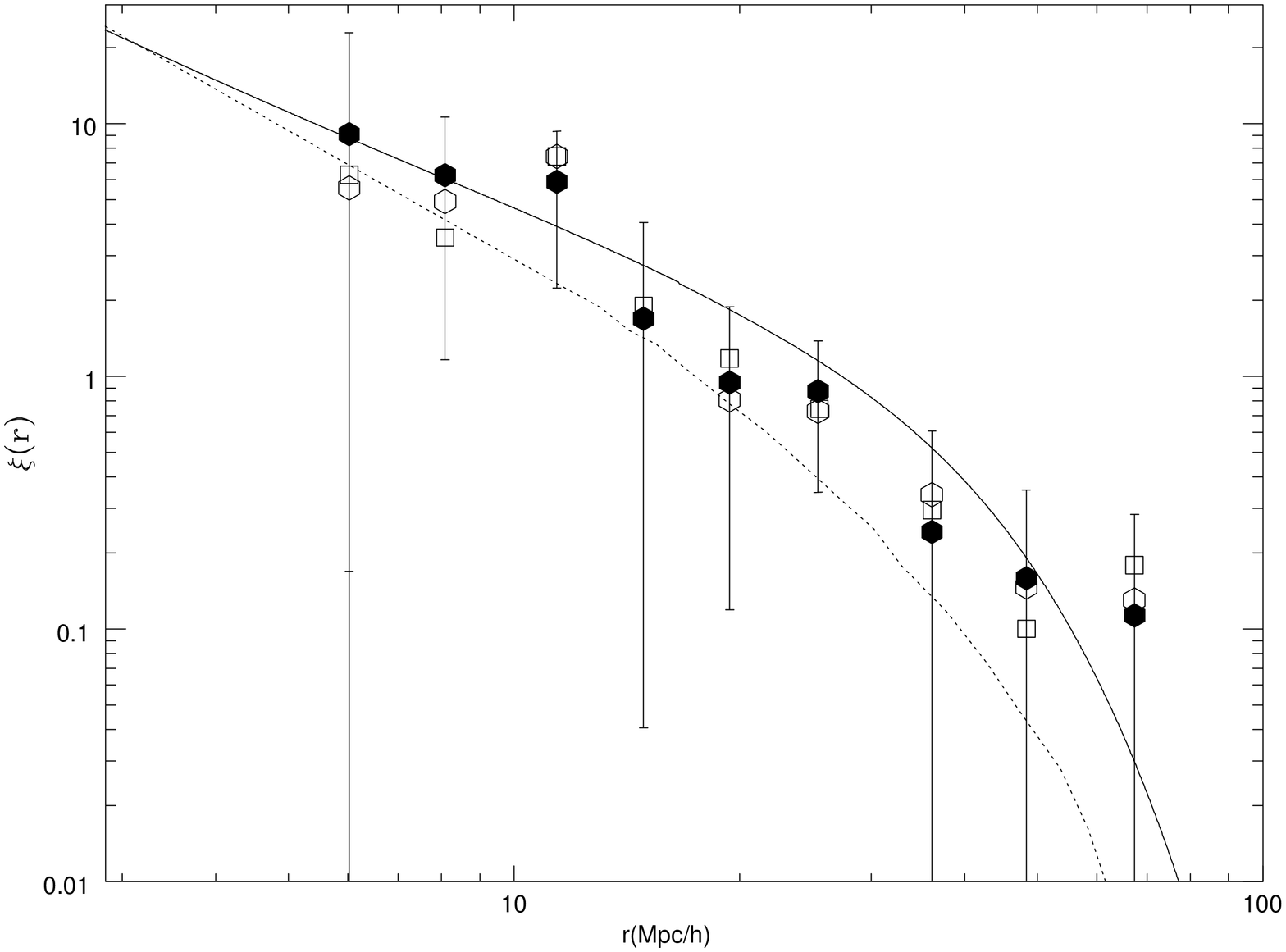,width=7.5cm,angle=-90} Fig. 5
\psfig{file=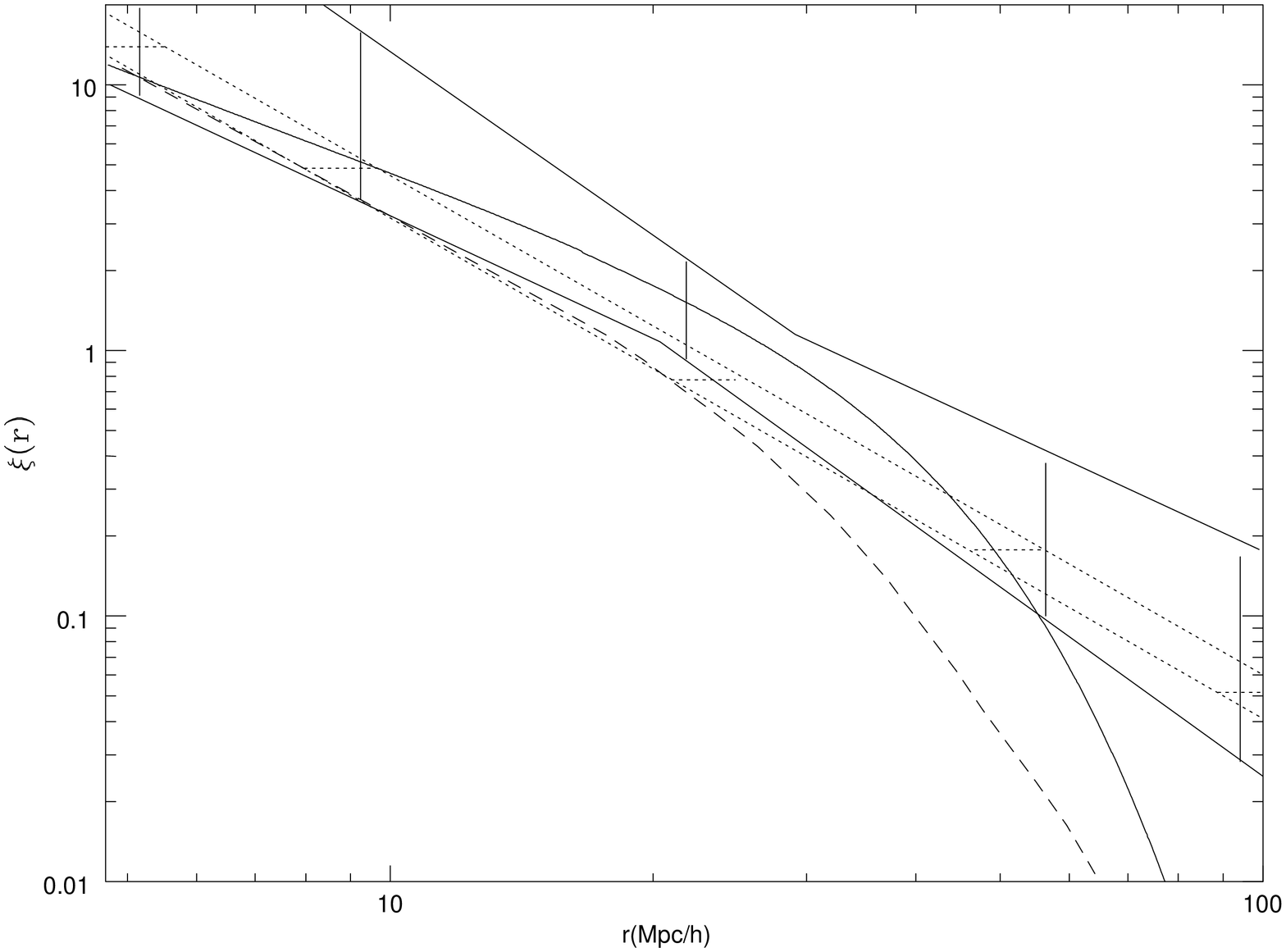,width=7.5cm,angle=-90} 
}} \centerline{\hbox{
}}
\caption{Comparison of the cluster space correlations in the XBACs sample
with the theoretical model of this paper. The observational estimates are
shown by two regions: the first (enclosed in the solid lines connected
by the vertical solid lines), refers to the (1$\protect\sigma $) estimates
obtained by Abadi, Lambas \& Muriel (1998), while the second region
(enclosed in the dashed lines connected by the horizontal dashed lines),
shows the (2$\protect\sigma $) estimates by Borgani, Plionis \&
Kolokotronis (1999). The solid curve represents the $\Omega _{\mathrm{0m}%
}=0.3$, $\protect\sigma _{8}=0.93$ $\Lambda $CDM model calculated using the
model of this paper, while the dashed line represents the prediction of
Moscardini et al. (2000a) for their $\Lambda $CDM model. }
\label{Fig. 1}
\end{figure}

\begin{figure}[tbp]
\centerline{\hbox{
\psfig{file=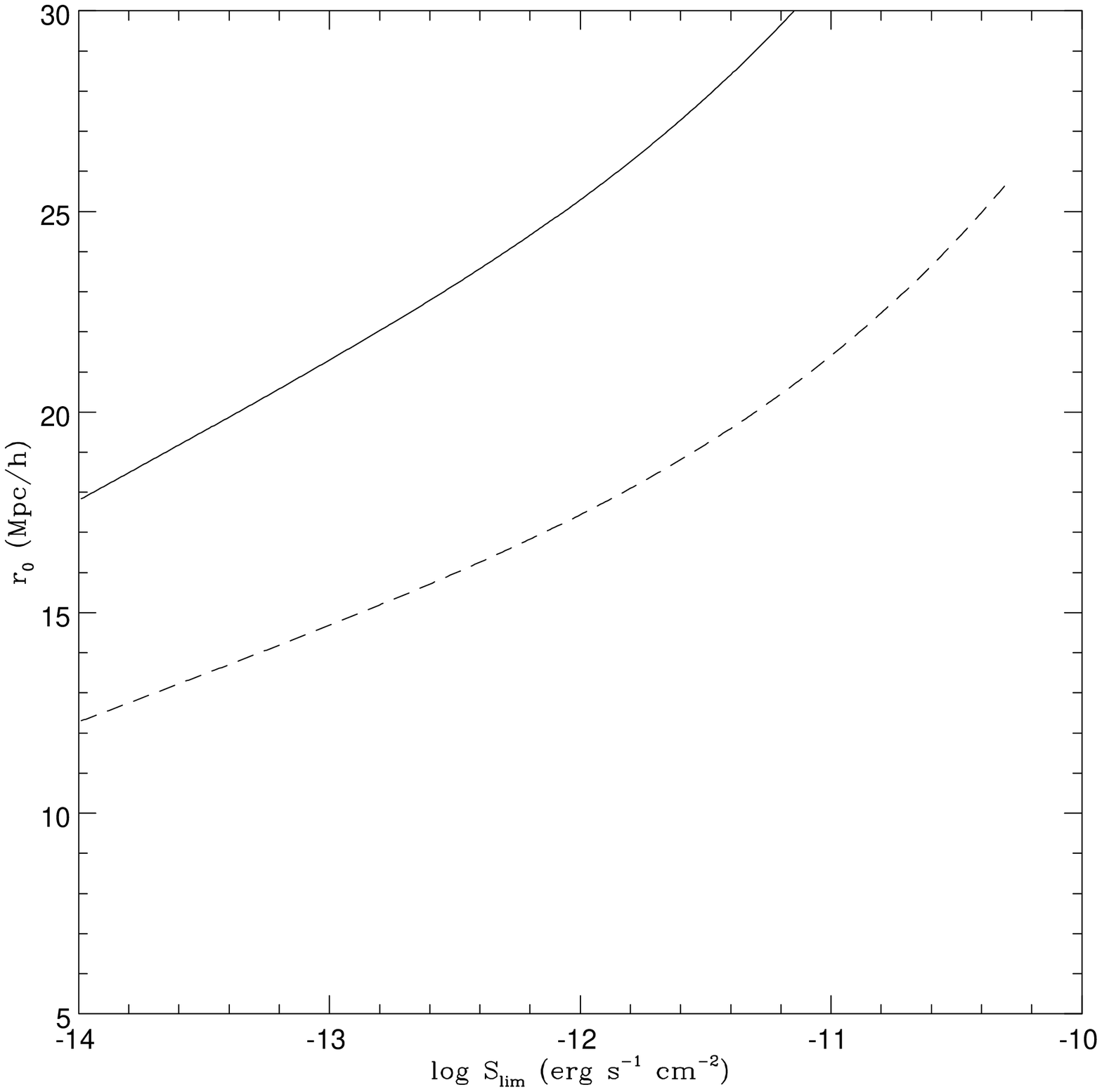,width=12cm}
}}
\caption{ The behavior of the correlation length $r_{0}$ as a function of
the limiting X-ray flux $S_{\mathrm{lim}}$. The solid line represents the $%
\Lambda $CDM model according to the model of this paper, and the dashed line
that of Moscardini et al. (2000a)}
\end{figure}

\begin{figure}[tbp]
\centerline{\hbox{
\psfig{file=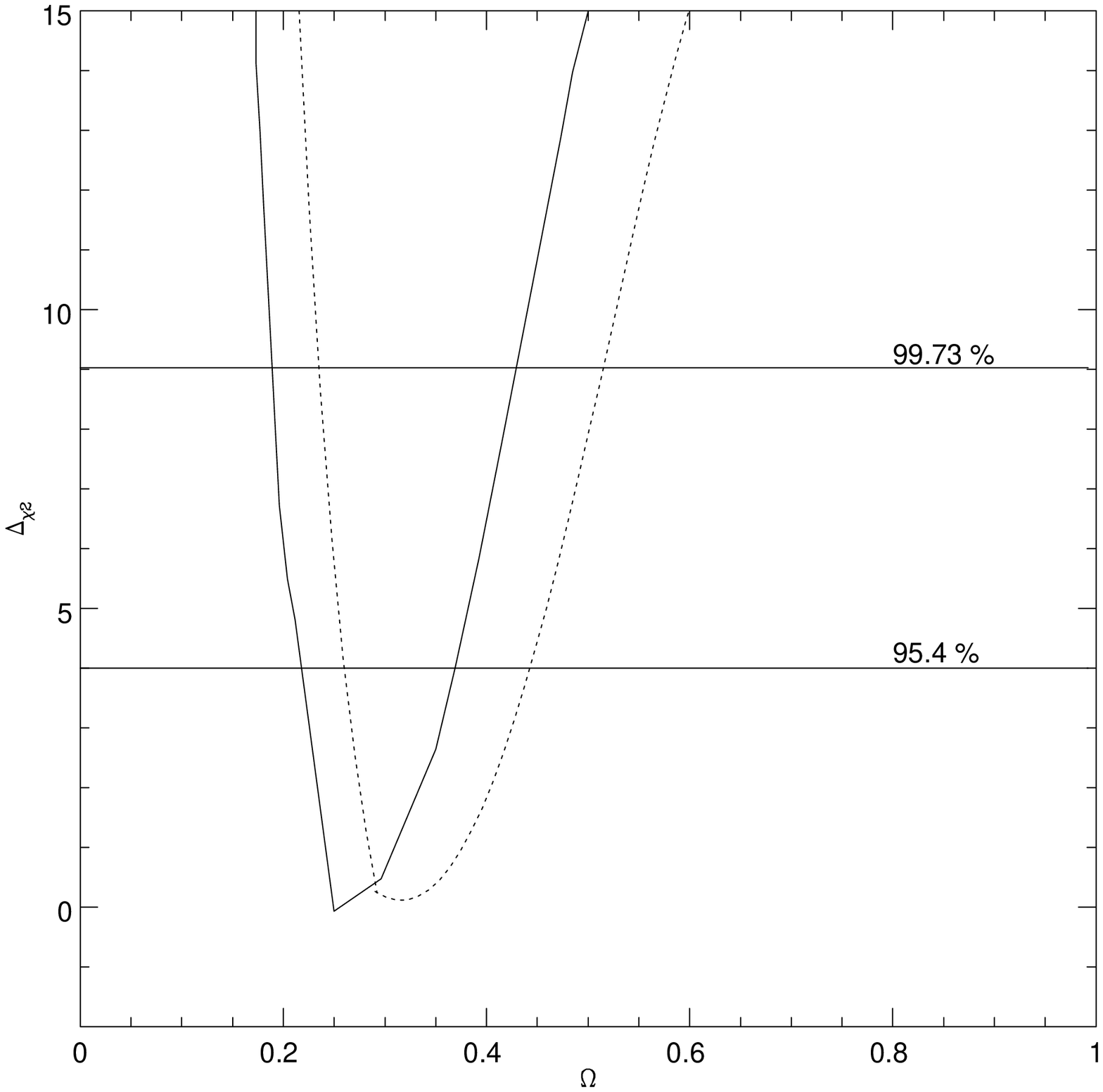,width=12cm}
}}
\caption{ The variation of $\Delta _{\protect\chi ^{2}}$ around the
best-fitting value of the matter density parameter $\Omega _{\mathrm{0m}}$
for flat CDM models 
(with varying cosmological constant $\Omega _{\mathrm{0\Lambda}}=1-\Omega _{\mathrm{0m}}$) with shape parameter $\Gamma =0.2$ and
normalization reproducing the cluster abundance. The solid line represents
the result of MMM (obtained using the X-ray complete cluster dataset), and
the dashed line that of this paper. Horizontal lines corresponding to the
95.4 and 99.73 per cent confidence levels are also shown.}
\end{figure}

\begin{figure}[tbp]
\caption{ Confidence contours of $\Gamma $ and $\Omega _{\mathrm{0m}}$.
Dashed and solid lines represent 95.4 and 99.73 per cent confidence levels.
The left panel refers to flat cosmological models with varying cosmological
constant the right one to open models with vanishing $\Omega _{\mathrm{%
0\Lambda }}$.}%
\centerline{\hbox{(a)
\psfig{file=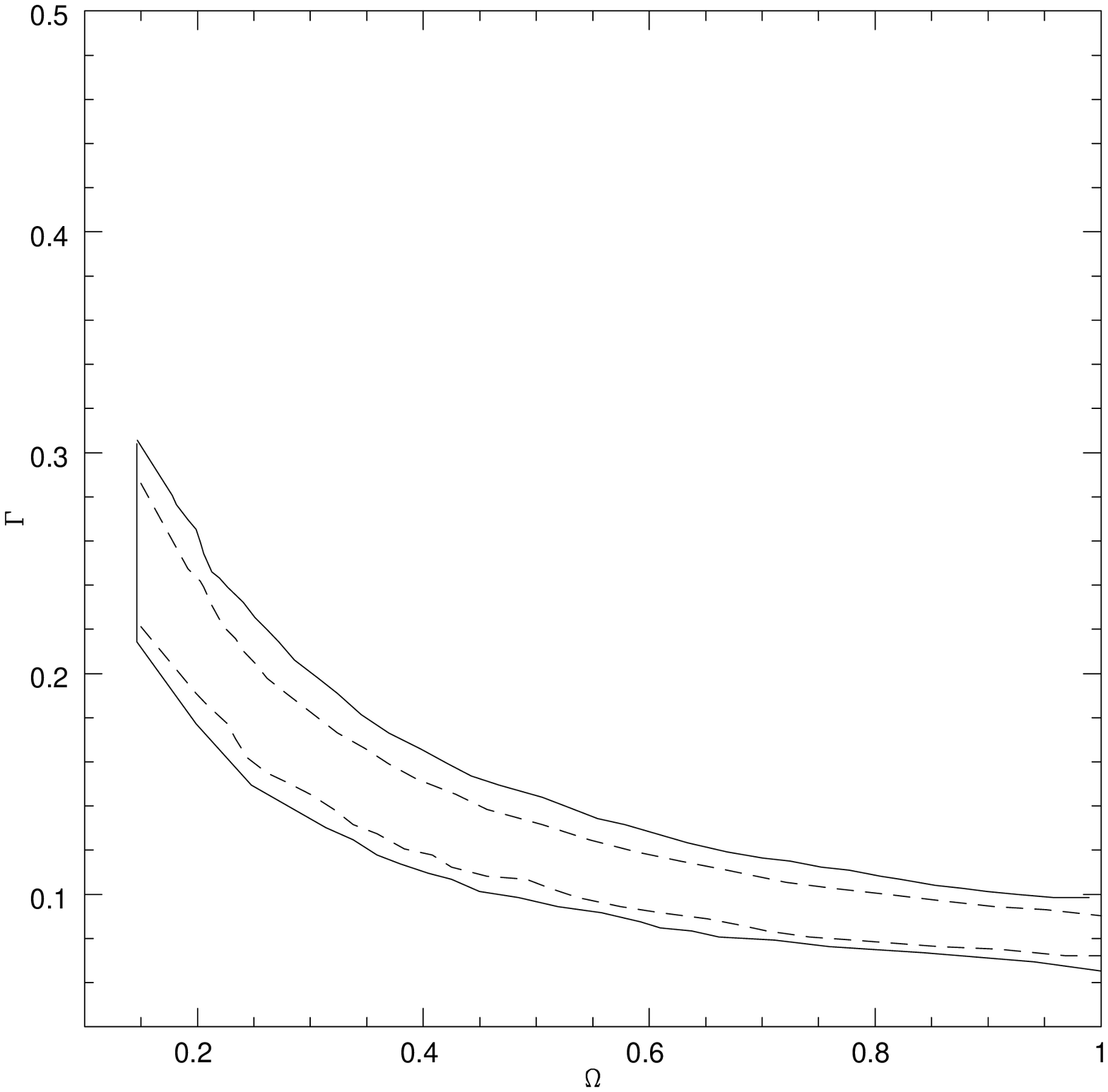,width=8cm} (b)
\psfig{file=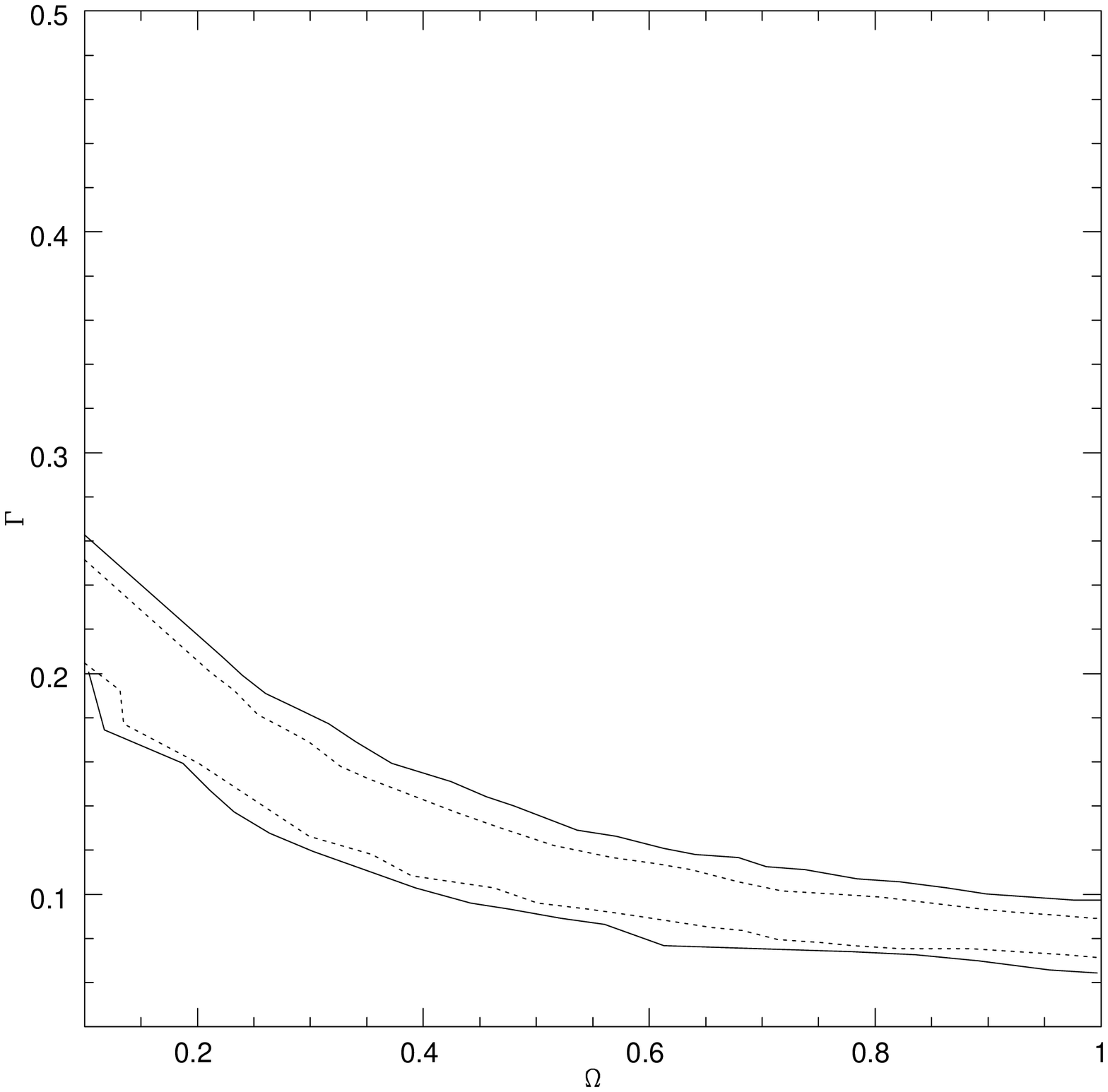,width=8cm}
}}
\par
\centerline{\hbox{
(a)
\hspace{1cm} (a)
\psfig{file=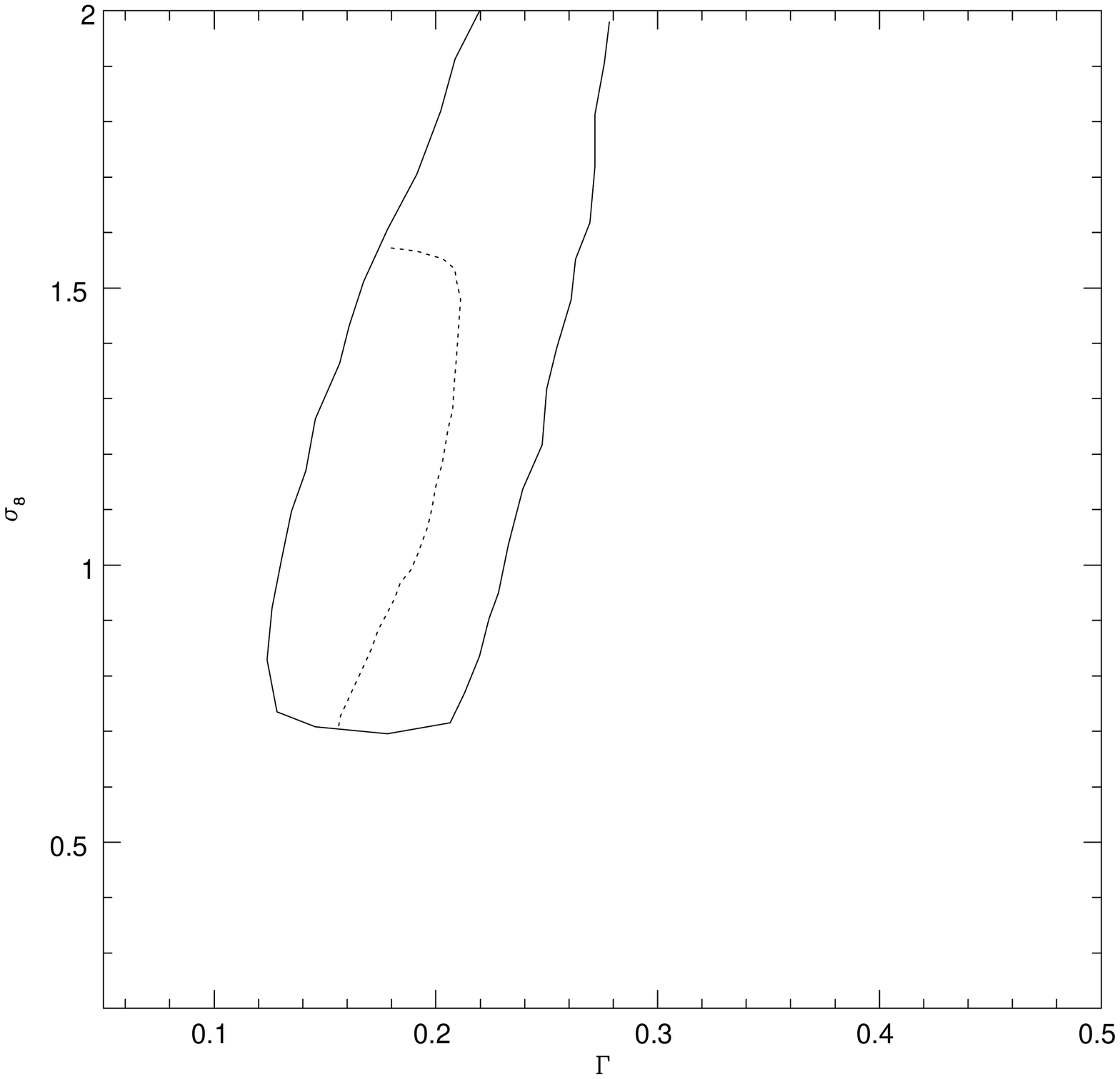,width=8cm}
\hspace{1cm} (b)
\psfig{file=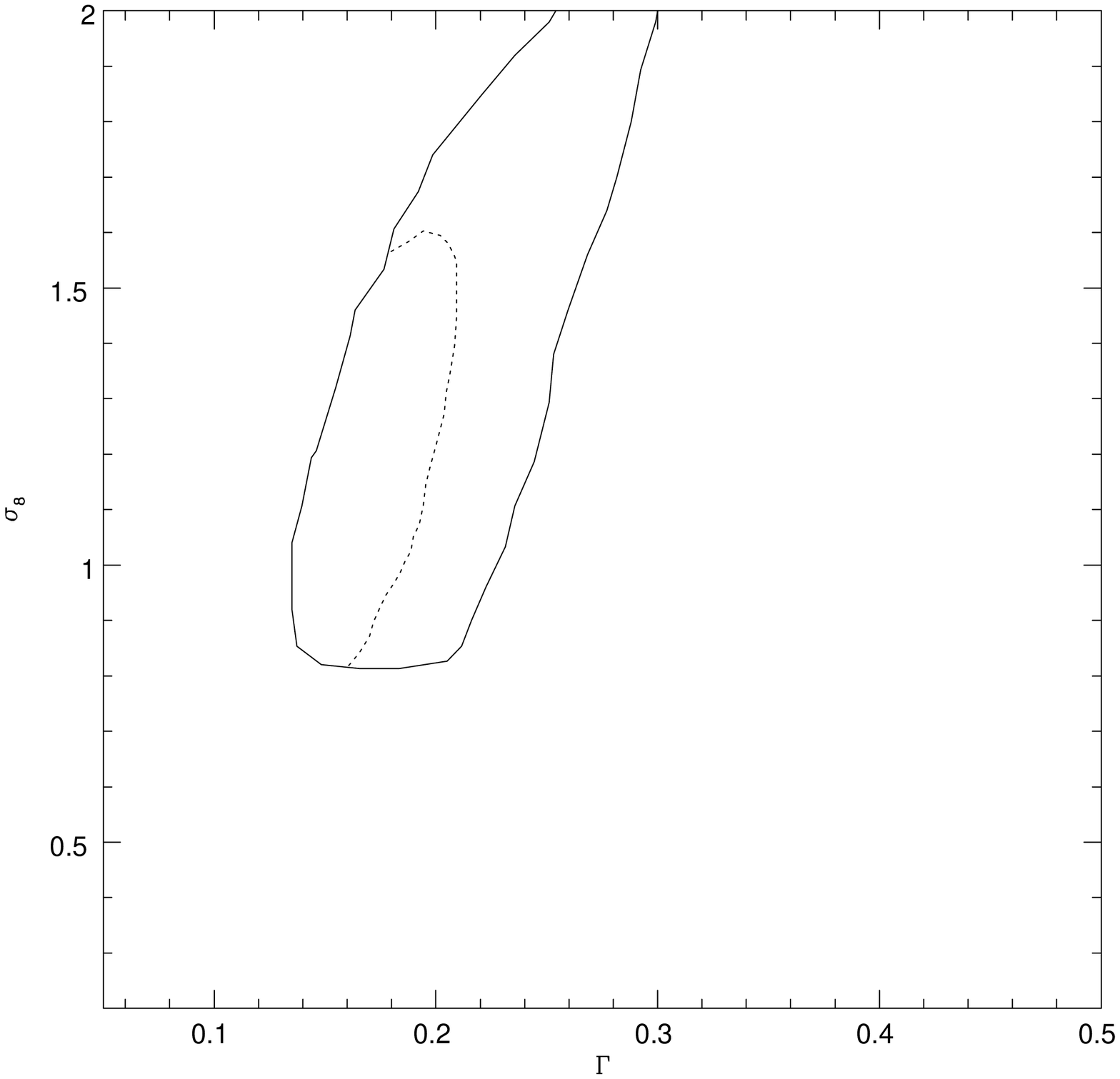,width=8cm}
}} 
\caption{Confidence contours of $\Gamma $ and $\protect\sigma _{8}$. Dotted
and solid lines represent 95.4 and 99.73 per cent confidence levels. Panel
(a) refers to a CDM model with $\Omega _{\mathrm{0m}}=0.3$, $\Omega _{%
\mathrm{0\Lambda }}=0$. Panel (b) refers to a CDM model with $\Omega _{%
\mathrm{0m}}=0.3$, $\Omega _{\mathrm{0\Lambda }}=0.7$.}
\label{FigM1gas}
\end{figure}


%

\section{Results}

We begin the analysis by comparing the theoretical predictions for the
two-point correlation function to the observational clustering properties of
RASS1 and afterwards to those of XBACs.

Fig. 4 shows the comparison with RASS1. In the plot, the dashed line
represents the prediction of Moscardini et al. (2000b) for their $\Lambda $%
CDM model, and the solid line represents the prediction of the $\Lambda $%
CDM, 
with $\Omega _{\mathrm{0m}}=0.3$, $\sigma _{8}=0.93$, and $\Gamma =0.21$
(see e.g. Liddle et al. 1996a,b and references therein), for the `default'
value of $\alpha =3.4$. 
The plot for the OCDM model was not reported since it gives very similar
predictions to the $\Lambda $CDM model. The $\Lambda $CDM model is in good
agreement with the data and the predictions are always inside the 1 $\sigma $
errorbars, $r_{0}\simeq 28.2 {\rm h^{-1}}$ Mpc. A comparison with the Moscardini
et al. (2000b) results, after the latter has been corrected taking account of
the description of clustering in the past-light cone, (which gives $%
r_{0}\simeq 22.4 {\rm h^{-1}}$ Mpc), 
shows that in our model the correlations are higher, 
by more than 20\%. 
Varying the value of $\alpha $ in the range 2.5-3.5, not plotted in Fig. 4, 
the resulting spatial correlation function does not change its shape, but
only its amplitude: the smaller $\alpha \ $is, the smaller $\xi (r)$ is.\
However, the changes are quite small, as shown in Fig. 11 of Moscardini et
al. (2000a). 
The induced change of $r_{0}$, can be written as $r_{0}\simeq 28.2\pm
5.2h^{-1}$ Mpc.

In Fig. 5 the model is compared with the XBACs catalogue. 
Abadi, Lambas \& Muriel (1998) found that the XBACs spatial correlation
function can be fitted by the usual power-law relation $\xi
(r)=(r/r_{0})^{-\gamma }$ with $\gamma =1.92$ and $r_{0}=21.1_{-2.3}^{+1.6}\
{\rm h^{-1}}$ Mpc (the errors correspond to 1 $\sigma $), while Borgani, Plionis
\& Kolokotronis (1999), who adopted an analytical approximation to the
bootstrap errors for the variance of $\xi (r)$, found $\gamma
=1.98_{-0.53}^{+0.35}$ and a slightly larger value of $%
r_{0}=26.0_{-4.7}^{+4.1}\ h^{-1}$ Mpc (the errors in this case are 2-$\sigma 
$ uncertainties). %
%
In Fig. 5, these observational estimates are compared with the theoretical
predictions of this paper. 
The observational estimates are shown by two regions: the first one
(enclosed in the solid lines connected by the vertical solid lines), refers
to the (1 $\sigma $) estimates obtained by Abadi, Lambas \& Muriel (1998),
while the second region (enclosed in the dashed lines connected by the
horizontal dashed lines), 
shows the (2 $\sigma $) estimates by Borgani, Plionis \& Kolokotronis
(1999). The solid curve represents the $\Omega _{\mathrm{0m}}=0.3$, $\sigma
_{8}=0.93$ $\Lambda $CDM model, while the dashed line represents the
prediction of Moscardini et al. (2000a) for their $\Lambda $CDM model 
\footnote{%
For the $\Lambda $CDM model (solid line), the results are in good agreement
with the observational data ($r_{0}\simeq 28.2 {\rm h^{-1}}$ Mpc).}. 
%
%

The results are in qualitative agreement with MMM, but the value for $r_{0}$
($r_{0}\simeq 28.2 {\rm h^{-1}}$ Mpc) obtained here is larger than the value in
the quoted paper 
($r_{0}\simeq 22h^{-1}$ Mpc for $\Lambda $CDM), while it is in better
agreement with that of Borgani, Plionis \& Kolokotronis (1999), who found a
value of $r_{0}=26_{-4.7}^{+4.1}{\rm h}^{-1}$ Mpc. Similar considerations to those
relative to Fig. 4 are valid if we vary the value of $\alpha $.

In order to study the possible dependence of the clustering properties of
the X-ray clusters on the observational characteristics defining the survey,
we plot the values of the correlation length $r_{0}$ in the catalogues where
we vary the limiting X-ray flux $S_{\mathrm{lim}}$. 
%
%
The result is shown in Fig. 6: the dashed line is the result obtained by
MMM, and the solid line that using the model of this paper, in the case of a 
$\Lambda $CDM model. The other cosmological models, not plotted, have a
similar behavior but smaller amplitude (see MMM). In Fig. 6, 
$r_{0}$ increases with $L_{\mathrm{lim}}$. The increase is more rapid for $%
\Lambda $CDM models (see MMM). A similar analysis has been made by S2000
(see their Fig. 8) and MMM. 
Even if \ the S2000 results cannot be directly compared with those of MMM,
(differences in cosmological parameters, formalism for the past-light cone
effect, etc.) or even with those of this paper (differences in cosmological
models parameters adopted), it is clear that there is a qualitative
agreement but MMM tends to predict smaller correlation lengths (by
approximately 20-30 \%) with respect to S2000 and the latter predicts
smaller correlation lengths than this paper. Part of the difference between
MMM and S2000 comes from the different values of the exponent of the
temperature used by the quoted authors in the temperature-luminosity
relation: in MMM $\mathcal{B}=1/3$ (see Sect. 2.4)), and that used in S2000 (%
$\alpha =1/3.4$ (see their Eq. 11)). 
Moreover, in their approach S2000 include a method to take account of
redshift-space distortion effects not completely considered by MMM 
\footnote{They used a zero-order model similar to that of Kaiser} which tends to
increase the correlation estimates.

In the present paper, the correlation lengths are larger in comparison with
both the previous two quoted papers 
predictions. These differences, expecially when we compare the results with
those of S2000, are due essentially to the different M-T relation, mass
function and bias model used in this paper. As reported in Sect. (2), the
fundamental goal of this paper is to study the effects of the improvements
quoted above on the values of the, cosmological parameters, and then 
to constrain the cosmological parameters by the clustering properties of
clusters of galaxy 
using a maximum-likelihood analysis. This analysis was started by
considering as a free parameter $\Omega _{m}$ only, fixing $\Gamma =0.2$,
which is in the range suggested by various other works (see e.g. Peacock \&
Dodds 1996), and $\sigma _{8}$ to reproduce the cluster abundance.

For the normalization we adopt the fitting formula by Pierpaoli et al.
(2003), 
%
%
with the M-T normalization parameter $T_{\ast}=1.75$ (see Pierpaoli et al.
2003). 

In Fig. 7, we plot the results of the maximum likelihood analysis obtained
by using X-ray band data (RASS1, BCS, XBACs, REFLEX) for flat models with
varying cosmological constant ($\Omega _{\mathrm{0\Lambda }}=1-\Omega _{%
\mathrm{0m}}$), with $\Gamma =0.2$. The solid line represents the results
obtained using the complete X-ray data set by MMM, and the dotted line gives
the results for the model in the present paper.

The constraints on $\Omega _{\mathrm{m}}$ are: $0.25\leq \Omega _{\mathrm{0m}%
}\leq 0.45$ and $0.23\leq \Omega _{\mathrm{0m}}\leq 0.52$ at 95.4 and 99.73
per cent levels, respectively. 
The result shows larger values for $\Omega _{\mathrm{0m}}$ on the order of
20-30 \%, when compared with MMM results ($0.2\leq \Omega _{\mathrm{0m}}\leq
0.35$ and $0.2\leq \Omega _{\mathrm{0m}}\leq 0.45$ at the 95.4 and 99.73 per
cent levels, respectively). 

In Fig. 8 we show the results of the maximum likelihood analysis fixing the
model normalization to reproduce the cluster abundance, and leaving two free
parameters: $\Gamma $, and $\Omega _{\mathrm{m0}}$. The solid lines
represents the 99.73 confidence levels, and the dashed lines the 95.4
confidence levels. 
%
%
%
The figure shows that the allowed regions, at least for $\Omega _{\mathrm{0m}%
}\geq 0.5$, depend strongly on $\Gamma $. In the case of the flat model
(Fig. (8a)), the constraints for $\Gamma $ are $0.1\leq \Gamma \leq 0.14$,
and in the case of the open model (Fig. (8b)) $0.09\leq \Gamma \leq 0.13$.
These values are smaller than those of MMM, by about $20-30\%$. (Note that
in MMM paper, the reported values are approximated. A comparison of our and
their confidence contours shows the difference).

%
%

In Fig. 9 we show the constraints in the $\Gamma -\sigma _{8}$ plane (after
keeping the values of $\Omega _{\Lambda }$ fixed). 
We consider an open model with $\Omega _{\mathrm{0m}}=0.3$, Fig. (9a), and a
flat model, Fig. (9b), again with $\Omega _{\mathrm{0m}}=0.3$. \footnote{%
Note that the value of $\Omega _{\mathrm{0m}}$ chosen is not too different
from that obtained by WMAP: $\Omega _{\mathrm{0m}}{\rm h}^{2}=0.14\pm 0.02$ and $%
{\rm h}=0.72\pm 0.05$ (Spergel et al. 2003)} 


For the open model with $\Omega _{\mathrm{0m}}=0.3$ the $2\sigma $ region
has $\Gamma $ in the range 0.11-0.2 and $\sigma _{8}$ between 0.7 and 1.55.
The values of $\sigma _{8}$ obtained here are larger by $\simeq 20\%$ than
those of MMM.

For a flat model with $\Omega _{\mathrm{0m}}=0.3$ the $2\sigma $ region has $%
0.13\leq \Gamma \leq 0.2$ and $0.8\leq \sigma _{8}\leq 1.6$. The values of $%
\sigma _{8}$ obtained are larger than those of MMM by $\simeq 20\%$

The effect of varying $\alpha $ on the maximum-likelihood analysis has been
also studied. From the study, it turns out that $\Gamma $ is quite
insensitive to a change of $\alpha $, while $\sigma _{8}$ is only weakly
dependent on it: the minimum in the maximum-likelihood analysis decreases
(increases) by $\simeq 10\%$ for $\alpha =2.5$, (for $\alpha =3.5$) 
(see also Fig. 8 of Borgani, Plionis \& Kolokotronis 1999). 
For $\Omega _{\mathrm{0m}}$ 
the changes are similar to those of $\Gamma $ but larger (of the order of 20
\%) 
(see also Fig. 4 of Borgani et al. 2001, and Fig. 10 of Eke et al. 1998).


Notice that in the analysis of this paper, we used only the data from the
X-ray catalogues and not a combination of optical and X-ray catalogues. This
is because, as shown by MMM, the combination of optical and X-ray catalogues
gives results that are almost indistinguishable from those obtained by the
X-ray analysis only. As in MMM, 
the constraints from the X-ray datasets are in general tighter than those
obtained from the optical data.

Although the quoted uncertainties have so far been of minor importance
because of the paucity of observational data, a breakthrough is needed in
the quality of the theoretical modeling if high-redshift clusters are to
take part in the high-precision-era of observational cosmology. As shown,
using models like PS for the mass function instead of models that are in
agreement with Jenkins et al. (2001), introduces errors of the order of
20-30 \% in the values of the parameters constrained (see Del Popolo 2003)
and other errors are introduced if one uses a simplified version of the M-T
relation. 
Moreover a proper interpretation of such redshift surveys in terms of the
clustering evolution requires an understanding of many cosmological effects
which can be neglected for $z<<1$ and thus have not been considered
seriously so far. For example, neglecting the light-cone effect leads to
underestimates of $\simeq 20\%$ in samples like RASS1, and of up to 25 \%
for a deeper survey such ABRIXAS (Moscardini et al. 2000a). Neglecting
red-shift space distortions produces underestimates of $r_{0}$ of $\simeq
10\%$: as reported in the present paper, a comparison between the behavior
of the correlation length as a function of the limiting X-ray flux, in
papers like S2000 and MMM shows a difference of $\simeq 30\%$ because in MMM
the redshift-space distortion effects were not taken into account, and
because of a difference in the exponent in the temperature-luminosity
relation. Taking into account the asphericity in gravitational collapse that
leads to a different relation for bias (Del Popolo \& Gambera 1998; Sheth \&
Tormen 1999), to a modified version of the mass function (Del Popolo \&
Gambera 1998; Sheth \& Tormen 2002; Del Popolo 2002b), and a different M-T
relation (Del Popolo 2002a), leads to higher values of $\Omega _{\mathrm{0m}%
} $, $\sigma _{8}$, $r_{0}$ by at least 20 \%. 
%
In the near future theory and observations should converge towards a more
precise constraining of cosmological parameters.

%
%

%
%

%
%
%


%

\section{Discussion and Conclusions}

In this paper, we have recalculated the two-point correlation function of
clusters of galaxies for OCDM and $\Lambda $CDM cosmological models,
improving the model of S2000, in the light of recent theoretical
developments, by using the theoretical mass function derived in Del Popolo
(2002a), the M-T relation derived in Del Popolo (2002b) and the bias model
of Del Popolo (2001). 
As in Suto's paper, the model properly takes account of nonlinear
gravitational evolution of mass fluctuations, redshift-space distortion due
to the linear peculiar velocity field and to finger-of-god, cluster
abundance and bias evolution, and the light-cone effect. 
This theoretical model has before been compared with the observed spatial
correlation function for clusters in RASS1, and in XBACs samples. The
comparison shows that only the predictions of models with $\Omega _{\mathrm{m%
}}=0.3$ are in good agreement with data. 
The results are in qualitative agreement with MMM, but the values for $r_{0}$
here obtained ($r_{0}\simeq 28.2 {\rm h^{-1}}$ Mpc for the $\Lambda $CDM model) are
larger than the values of the quoted paper (MMM), 
($r_{0}\simeq 22 {\rm h^{-1}}$ Mpc for $\Lambda $CDM), while they are in better
agreement with that of Borgani, Plionis \& Kolokotronis (1999), who found a
value of $r_{0}=26_{-4.7}^{+4.1}{\rm h}^{-1}$ Mpc. In order to study the possible
dependence of the clustering properties of the X-ray clusters on the
observational characteristics defining the survey, we plot the values of the
correlation length $r_{0}$ in the catalogues where we vary the limiting
X-ray flux $S_{\mathrm{lim}}$. 
%
\footnote{%
Notice that this analysis can be related to the study of the richness
dependence of the cluster correlation function. In fact, a change in the
observational limits implies a change in the expected mean intercluster
separation $d_{c}$.} All the cosmological models displays an increase of $%
r_{0}$ with $L_{\mathrm{lim}}$. The increase is more rapid for $\Lambda $CDM
models (see MMM). Comparing the result with those of a similar analysis by
S2000, (see their Fig. 8) and MMM it is clear that there is a qualitative
agreement but MMM tends to predict smaller correlation lengths (by
approximately 20-30 \%) with respect to S2000, and the latter predicts
smaller correlation lengths than the present paper. 
These differences, expecially when we compare the results with those of
S2000, are due essentially to the different M-T relation and mass function
used in this paper. In order to obtain constraints on cosmological
parameters 
we performed a maximum-likelihood analysis by comparing the theoretical
predictions to a set of observational data in the X-ray band (RASS1 Bright
Sample, BCS, XBACs, REFLEX), similarly to MMM. The parameters to be
constrained are: $\Omega _{\mathrm{m}}$, $\Omega _{\Lambda }$, the
power-spectrum shape parameter $\Gamma $ and the normalization $\sigma _{8}$%
. The constraints obtained for the matter density parameter in a flat CDM
model, are: $0.25\leq \Omega _{\mathrm{0m}}\leq 0.45$ and $0.23\leq \Omega _{%
\mathrm{0m}}\leq 0.52$ at the 95.4 and 99.73 per cent levels, respectively,
larger by $\simeq 20\%$ than the MMM predictions. %
%
Keeping the model normalization fixed to reproduce the cluster abundance,
and leaving two free parameters: $\Gamma $, and $\Omega _{\mathrm{m0}}$, we
find that for the flat model the constraints for $\Gamma $ are $0.1\leq
\Gamma \leq 0.14$, while for the open model $0.09\leq \Gamma \leq 0.13$.
These values are smaller than those of MMM, by about $20-30\%$. After fixing
the values of $\Omega _{\Lambda }$, we obtain the constraints in the $\Gamma
-\sigma _{8}$ plane, showing that 
if we keep the value of $\Omega _{\Lambda }$ fixed the open model with $%
\Omega _{\mathrm{0m}}=0.3$ the $2\sigma $ region has $\Gamma $ in the range
0.11-0.2 and $\sigma _{8}$ between 0.7 and 1.55. 
In the flat model with $\Omega _{\mathrm{0m}}=0.3$ the $2\sigma $ region has 
$0.13\leq \Gamma \leq 0.2$ and $0.8\leq \sigma _{8}\leq 1.6$ In all three
cases, the values of $\sigma _{8}$ obtained are larger than those of MMM by $%
\simeq 20\%$. Varying $\alpha $, it turns out that $\Gamma $ is quite
insensitive to the change of $\alpha $, while $\sigma _{8}$ is only weakly
dependent on it: the minimum in the maximum-likelihood analysis decreases
(increases) by  $\simeq 10\%$ for $\alpha =2.5$, (for $\alpha =3.5$). In the
case of $\Omega _{\mathrm{0m}}$ the changes are similar to those of $\Gamma $
but larger (of the order of 20 \%). Allowing the shape parameter to vary, we
find that the clustering properties of clusters are almost independent of
the matter density parameter and of the presence of a cosmological constant,
while they appear to be strongly dependent on the shape parameter. The
constraints from X-ray data are tighter than those coming from optical data.
In conclusion, the data on clustering properties of galaxies can be used to
constrain important cosmological parameters like $\Omega _{\mathrm{0m}}$, $%
\Gamma $ and $\sigma _{8}$.

\section*{Acknowledgments}

A. Del Popolo would like to thank Bo$\breve{g}azi$\c{c}i University Research
Foundation for the financial support through the project code 01B304.



{}

\end{document}